\documentclass[final,authoryear,12pt]{elsarticle}
\usepackage{graphicx,subfigure}
\usepackage{longtable}

\biboptions{authoryear}

\journal{Planetary and Space Science}

\begin{document}

\hyphenation{EURONEAR}
\hyphenation{Vaduvescu}

\begin{frontmatter}



\title{739 observed NEAs and new 2-4m survey statistics within the EURONEAR network. \tnoteref{label1}}
\tnotetext[label1]{Based on observations taken with the following telescopes: Blanco 4m in Chile, INT 2.5m and WHT 4.2m in La Palma, 
OHP 1.2m and Pic du Midi 1m in France, Tautenburg 2m and Bonn AIfA 0.5m in Germany, Galati 0.4m and Urseanu 0.3m in Romania. }


\author[inst1,inst2,inst3]{O. Vaduvescu\corref{cor1}}
\ead{ovidiuv@ing.iac.es}

\author[inst2]{M.~Birlan}
\author[inst4,inst5]{A.~Tudorica}
\author[inst2,inst6,inst8]{M.~Popescu}
\author[inst2]{F.~Colas}
\author[inst10]{D.~J.~Asher} 

\author[inst7,inst8]{A.~Sonka}
\author[inst9,inst11]{O.~Suciu}
\author[inst12,inst13,inst16]{D.~Lacatus}
\author[inst12,inst13,inst16]{A.~Paraschiv}
\author[inst14]{T.~Badescu}

\author[inst15,inst16]{O.~Tercu}
\author[inst15,inst16]{A.~Dumitriu}
\author[inst15,inst16]{A.~Chirila}

\author[inst17]{B.~Stecklum}
\author[inst3,18]{J.~Licandro}

\author[inst2,inst6]{A.~Nedelcu}
\author[inst19]{E.~Turcu}
\author[inst2]{F.~Vachier}
\author[inst2]{L.~Beauvalet}
\author[inst2]{F.~Taris}
\author[inst2]{L.~Bouquillon}

\author[inst20]{F.~Pozo~Nunez}
\author[inst21]{J.~P.~Colque~Saavedra}
\author[inst21]{E.~Unda-Sanzana}

\author[inst1,inst22,inst23]{M.~Karami}
\author[inst22]{H.~G.~Khosroshahi}
\author[inst10,inst11]{R.~Toma} 
\author[inst1,inst26]{H.~Ledo}
\author[inst24,inst1]{A.~Tyndall}
\author[inst1,inst25]{L.~Patrick}
\author[inst1,inst27]{D.~F\"ohring}

\author[inst14]{D.~Muelheims}
\author[inst14]{G.~Enzian}
\author[inst14]{D.~Klaes}
\author[inst14]{D.~Lenz}
\author[inst14]{P.~Mahlberg}
\author[inst14]{Y.~Ordenes}
\author[inst14]{K.~Sendlinger}

\cortext[cor1]{Corresponding author}

\address[inst1]{\scriptsize  {Isaac Newton Group of Telescopes (ING), Apartado de Correos 321, E-38700 Santa Cruz de la Palma, Canary Islands, Spain}}
\address[inst2]{\scriptsize  {IMCCE, Observatoire de Paris, 77 Avenue Denfert-Rochereau, 75014 Paris Cedex, France}}
\address[inst3]{\scriptsize  {Instituto de Astrof\'isica de Canarias (IAC), C/V\'ia L\'actea s/n, 38205 La Laguna, Spain}}
\address[inst4]{\scriptsize  {Bonn Cologne Graduate School of Physics and Astronomy, Germany}}
\address[inst5]{\scriptsize  {Argelander-Institut f\"ur Astronomie, Universit\"{a}t Bonn, Auf dem H\"{u}gel 71, D-53121 Bonn, Germany}}
\address[inst6]{\scriptsize  {The Astronomical Institute of the Romanian Academy, Cutitul de Argint 5, 040557 Bucharest, Romania}}
\address[inst7]{\scriptsize  {The Astronomical Observatory ``Admiral Vasile Urseanu'', B-dul Lascar Catargiu 21, Bucharest, Romania}}
\address[inst8]{\scriptsize  {Bucharest Astroclub, B-dul Lascar Catargiu 21, sect 1, Bucharest, Romania}}
\address[inst9]{\scriptsize  {Babes-Bolyai University, Faculty of Mathematics and Informatics, 400084 Cluj-Napoca, Romania}}
\address[inst10]{\scriptsize {Armagh Observatory, College Hill, Armagh BT61 9DG, Northern Ireland, United Kingdom}}
\address[inst11]{\scriptsize {Romanian Society for Meteors and Astronomy (SARM), CP 14 OP 1, 130170, Targoviste, Romania}}
\address[inst12]{\scriptsize {Research Center for Atomic Physics and Astrophysics, Faculty of Physics, University of Bucharest, Atomistilor 405, CP Mg-11, 077125 Magurele, Ilfov, Romania}}
\address[inst13]{\scriptsize {Institute of Geodynamics Sabba S. Stefanescu, Jean-Louis Calderon 19-21, Bucharest, Romania, RO-020032, Romania}}
\address[inst14]{\scriptsize {Rheinische-Friedrich-Wilhelms Universitaet Bonn, Argelander-Institut fur Astronomie, Auf dem Hugel 71 D-53121 Bonn, Germany}}

\address[inst15]{\scriptsize {Galati Astronomical Observatory of the Natural Sciences Museum Complex, Str. Regiment 11-Siret, no 6A, 800340, Galati, Romania}}
\address[inst16]{\scriptsize {Calin Popovici Astronomy Club, Str. Regiment 11-Siret, no 6A, 800340, Galati, Romania}}

\address[inst17]{\scriptsize {T\"uringer Landessternwarte Tautenburg, Sternwarte 5, D - 07778 Tautenburg, Germany}}
\address[inst18]{\scriptsize {Departamento de Astrof\'isica, Universidad de La Laguna, E-38205 La Laguna, Tenerife, Spain}}
\address[inst19]{\scriptsize {The Astronomical Observatory of ``Stefan cel Mare'' University of Suceava, Str. Universitatii 13, 720229 Suceava, Romania}}
\address[inst20]{\scriptsize {Instituto de Astronomia, Universidad Catolica del Norte, Avenida Angamos 0610, Antofagasta, Chile}}
\address[inst21]{\scriptsize {Unidad de Astronomia, Universidad de Antofagasta, Avda. Angamos 601, 1270300, Antofagasta, Chile}}

\address[inst22]{\scriptsize {School of Astronomy, Institute for Research in Fundamental Sciences (IPM), PO Box 19395-5531,Tehran, Iran}}
\address[inst23]{\scriptsize {Sharif University of Technology, Azadi Ave.,  PO Box 11365-11155, Tehran, Iran}}
\address[inst24]{\scriptsize {Jodrell Bank Centre for Astrophysics, School of Physics and Astronomy, University of Manchester, M13 9PL, United Kingdom}}
\address[inst25]{\scriptsize {Department of Physics and Astronomy, University of Sheffield, Sheffield S3 7RH, United Kingdom}}
\address[inst26]{\scriptsize {Centre for Astrophysics Research, University of Hertfordshire, College Lane, Hatfield AL10 9AB, United Kingdom}}
\address[inst27]{\scriptsize {Department of Physics, Centre for Advanced Instrumentation, University of Durham, South Road, Durham, DH1 3LE, United Kingdom}}

\begin{abstract}
\small

We report follow-up observations of 477 program Near-Earth Asteroids (NEAs)
using nine telescopes of the EURONEAR network having apertures between 0.3
and 4.2 m.  Adding these NEAs to our previous results we now count 739
program NEAs followed-up by the EURONEAR network since 2006.  The targets
were selected using EURONEAR planning tools focusing on high priority
objects.  Analyzing the resulting orbital improvements suggests astrometric
follow-up is most important days to weeks after discovery, with recovery 
at a new opposition also valuable. 
Additionally we observed 40 survey fields spanning three nights covering
11 square degrees near opposition, using the Wide Field Camera on the 2.5m 
Isaac Newton Telescope (INT), resulting in 104 discovered main belt asteroids 
(MBAs) and another 626 unknown one-night objects. These fields, plus program NEA
fields from the INT and from the wide field MOSAIC II camera on the Blanco 4m 
telescope, generated around 12,000 observations of 2,000 minor planets 
(mostly MBAs) observed in 34 square degrees. We identify Near Earth Object (NEO) 
candidates among the unknown (single night) objects using three selection criteria.  
Testing these criteria on the (known) program NEAs shows the best selection 
methods are our $\epsilon-\mu$ model which checks solar elongation and 
sky motion and the MPC's NEO rating tool. 
Our new data show that on average 0.5 NEO candidates per square degree should be 
observable in a 2m-class survey (in agreement with past results), while an 
average of 2.7 NEO candidates per square degree should be observable in a 4m-class 
survey (although our Blanco statistics were affected by clouds). At opposition 
just over 100 MBAs (1.6 unknown to every 1 known) per square degree are detectable 
to $R=22$ in a 2m survey based on the INT data (in accordance with other results), 
while our two best ecliptic Blanco fields away from opposition lead to 135 MBAs 
(2 unknown to every 1 known) to $R=23$. 

% **** Maybe long enough already; omit this sentence:
% Though observing 2 mag fainter, orbit estimates suggest Blanco picks up
% proportionally more MBAs at brighter absolute magnitude, compared to INT,
% because Blanco is seeing further into the main belt.

\end{abstract}


\end{frontmatter}

% \linenumbers

%% ==================================================================================================

\section{Introduction}
\label{intro}

Since Ceres was found by Piazzi in 1801, the discovery rate of Main Belt
Asteroids (MBAs) has increased, to the point where the IAU Minor Planet
Center (MPC) has now cataloged over 600,000 minor planets \citep{mpc12a}.
Near Earth Asteroids (NEAs), defined as minor planets with perihelion
distance $q\leq1.3$ AU and aphelion distance $Q\geq0.983$ AU \citep{mor02},
represent an important Solar System population: their main formation
mechanisms include migration of MBAs due to resonances, especially 3:1
mean-motion with Jupiter and $\nu_6$ with Saturn \citep{far93}, possibly
combined with the Yarkovsky/YORP effects \citep{bot06}.  There are more than
9,200 known NEAs today \citep{mpc12a}.

Potentially Hazardous Asteroids (PHAs) are defined as a sub-class of NEAs 
having minimum orbital intersection distance $MOID\leq0.05$ AU and absolute 
magnitude $H\leq22$ \citep{bm94}. We know today more than 1,300 PHAs \citep{mpc12a}. 
Virtual Impactors (VIs) are NEAs whose future Earth impact probability is
non-zero according to the actual orbital uncertainty \citep{mg09}. There are
about 420 and 350 VIs listed in the NASA JPL Sentry Risk Table \citep{jpl12}
and NEODyS Risk List \citep{neo12}, respectively.

There is a continual need to observe asteroids in order to study their orbits
and, for NEAs, to catalog future approaches to Earth. New astrometry 
improves the orbital quality which would otherwise be degraded by effects 
including close planetary approaches and non-gravitational effects
(Yarkovsky, YORP).

The European Near Earth Asteroid Research (EURONEAR) was initiated in 
2005 in Paris in order to bring some European contribution to the NEA research 
using existing telescopes and hopefully some automated dedicated facilities 
\citep{eur12a}. Lacking dedicated funding, during recent years mostly 
volunteering students and amateur astronomers directed by a few researchers 
reduced the data in near real time (few hours or days after acquisition), 
some of them participating in observing runs. 
\cite{vad08} introduced EURONEAR and first observations obtained at Pic du
Midi in 2006, followed by \cite{bir10a} who described the first 160 observed
NEAs using nine telescopes available to the EURONEAR network during the first
four years. \cite{vad11} presented some MBA and NEA statistics based on three
runs using large field 1-2m facilities (Swope 1m, ESO/MPG 2.2m and INT 2.5m).

In this paper we present new NEA recovery and follow-up observations using
six professional and three educational-amateur telescopes available to the
EURONEAR network during the last two years.  Adding these observations, the
total number of observed NEAs within the EURONEAR network reaches 739 objects
(Oct 2012).  Thanks to two observing runs and a few discretionary time hours
awarded to use the large field imaging cameras of the 2.5m Isaac Newton
Telescope (INT) and Blanco 4m telescope, in a team comprised mostly of
students and amateur astronomers we obtained and carefully analyzed 572 CCD
fields covering 34 square degrees, visually scanning, measuring and reporting
around 12 thousand positions of more than two thousand asteroids.  We have
then used the data obtained with 2-4m telescopes to derive MBA and NEA
statistics going beyond our previous results based on data obtained with 1-2m
facilities \citep{vad11}.

Section~\ref{obsred} describes our observations. In Section~\ref{results} we
compile the results, including the classification into observed NEAs, known
MBAs and unknown objects. In Section~\ref{disc} we discuss these results,
focusing on the distribution of the known and unknown MBAs and NEAs; we
compare the INT and Blanco facilities and present statistics for the use of
2m and 4m class surveys. Conclusions and two future projects are presented in
Section~\ref{future}.

%% ==========================================================================

\section{Observations and data reduction}
\label{obsred}

To distinguish between targeted and new candidate NEAs, we define as {\it program NEA} any Near 
Earth Asteroid (NEA, PHA or VI) programmed for follow-up within the EURONEAR network. To plan our 
runs and select program NEAs given an observing place and date, we used two planning tools 
\citep{eur12b} which target the newly discovered objects from the Spaceguard priority list 
\citep{spa12} and the NEA bright and faint recovery opportunity lists maintained by the 
Minor Planet Center \citep{mpc12a}. More information about these planning tools can be found 
in \cite{vad11}. 

Ten cameras mounted on nine telescopes were used between Jan 2009 and June 2012. 
In Table~\ref{table1} we list specifications of each facility. We present each observing node 
and results next. 
For all observing sites and runs except TLS we used Astrometrica \citep{raa12} to identify 
program NEAs and other moving sources, using its field recognition, image registering and blinking 
capabilities, by visually scanning, measuring and reporting the observed fields in near real 
time (few hours or days after the runs). For relatively sparse observations and small sky 
area covered, Astrometrica has been proven an excellent team capability for use by students 
and amateurs. Moreover, for limited amounts of data, the human eye and brain have been better 
for moving object detection, reaching lower S/N detection versus automated pipelines (see 
for example \cite{vad11} for some references and comparison).

\newcommand {\0} {\phantom{0}}

\begin{table*}
\caption{Telescopes and detectors used for follow-up, recovery and securing
  orbits of NEAs. The columns give the observatory, telescope aperture (m), CCD camera, 
  number of CCDs and individual size in pixels, field of view and pixel scale.}
\label{table1}
\begin{center}
\resizebox{15cm}{!}{
\begin{tabular}{lclccc}
\hline
\hline \\
Telescope         & Aperture & Camera        & CCDs \& Pixels             & FOV ($^\prime$)   & Scale ($^{\prime\prime}$/pix) \\ 
\\
\hline 
\\
ORM INT           & 2.5    & WFC             &  $4\times(2048\times4096)$ & $34\times34$ L    & 0.333 \\ 
ORM WHT           & 4.2    & ACAM            &  $2048\times4096$          & $8$ diam          & 0.254 \\ 
CTIO Blanco       & 4.0    & MOSAIC II       &  $8\times(2048\times4096)$ & $36\times36$      & 0.27\0 \\ 
Pic du Midi (b)   & 1.05   & iKon-L Andor    &  $2048\times2048$          & $7.5\times7.5$    & 0.22\0 \\
Pic du Midi (c)   & 1.05   & Atik 383L+      &  $3326\times2504$          & $7.8\times5.8$    & 0.14\0 \\
Haute Provence    & 1.20   & TK1024          &  $1024\times1024$          & $11.7\times11.7$  & 0.685 \\
TLR Tautenburg    & 2/1.34 & SITe            &  $2048\times2048$          & $42\times42$      & 1.23\0 \\ 
Bonn/AIfA         & 0.50   & SBIG-STL 6303E  &  $3072\times2048$          & $23.0\times15.4$  & 0.45\0 \\
Galati            & 0.40   & SBIG STL-6303E  &  $3072\times2048$          & $29.8\times19.9$  & 0.58\0 \\
Urseanu           & 0.30   & TC273-home made &  $640\times500$            & $21.6\times16.9$  & 2.30\0 \\
\\
\hline
%\hline  
\end{tabular}
}
\end{center}
\end{table*}

\subsection{INT observing runs}

We used the 2.5m Isaac Newton Telescope (INT) owned by the Isaac Newton Group (ING) 
at Roque de Los Muchachos Observatory (ORM) in La Palma. Four nights were awarded by the Spanish 
Time Allocation Committee (CAT, 25--28 February 2012, proposal number C6, PI: O. Vaduvescu) and 
approximately 10 hours were used in total during four service/discretionary ING nights (S/D) 
in 2011. 

At the prime focus of the INT we used the Wide Field Camera (WFC) which consists of four CCDs 
$2K\times4K$ pixels each covering an L-shaped $34^\prime \times 34^\prime$ with a pixel scale 
of $0.33~^{\prime\prime}$/pix. All frames were observed with $2\times2$ binning 
($0.66~^{\prime\prime}$/pix) to decrease the readout time, except for the December 2011 run 
which was operated without binning. During three of the S/D nights (except for December 2011 run) 
we tracked the observed fields at half the NEA proper motion, while for the remaining runs we 
tracked fields at normal sidereal rate. The filling factor of the WFC (defined as the ratio of 
the active area over total sky subtended camera area) is 0.86. 
We used an $R$ filter for all runs\footnote{Historically we used $R$ filter for most of our 
EURONEAR runs in order to minimize moonlight, artificial light pollution and also for photometric 
consistency}. 

The CAT time was dark, and the weather was good during three nights, except for the first 
three hours during the second night and the third night when we closed due to humidity. 
The typical seeing at the INT was $\sim1.5^{\prime\prime}$.
For data reduction we used THELI \citep{erb05} to correct the field distortion in the 
prime focus of the INT, then using Astrometrica with a fit order 1-2 and USNO-B1 or NOMAD 
(thus UCAC2) catalogs. 

Besides the program NEAs (listed in Table A1 in the Appendix),
during three clear nights at the INT in February 2012 (the first, 
second and fourth of our run) we undertook an MBA mini-survey around opposition, covering in total 
40 WFC fields (five successive images of the same field) during 12 hours total time. The 
pointing grid of the MBA mini-survey was optimized in order to recover most new MBAs observed 
on previous night(s), assuming a daily apparent motion of $17^\prime$ westward
along the ecliptic (based on $0.7^{\prime\prime}$/min average speed). During the first night we 
observed 10 arbitrarily chosen WFC fields close to opposition in the ecliptic. During the second 
night we observed the same 10 WFC fields shifted by $-1$min in $\alpha$ and $+8^\prime$ 
in $\delta$. During the fourth night we observed a pair of fields shifted by $-2$min in 
$\alpha$ and $+16^\prime$ and $-1^\prime$ in $\delta$ with respect to
the second night. During the short time available for this mini-survey, 
this pointing grid allowed us to recover many objects appearing in previous nights 
in the WFC fields and maximize the number of MBA discoveries. 

Due to the large WFC field and INT aperture, we measured and identified all moving sources 
detected visually via blinking in all observed fields in all INT runs, reporting all 
known and new objects visible up to apparent magnitude $R\sim22$. In total, at the INT we 
observed 87 WFC fields covering $\sim24$ square degrees (including 40 WFC fields covering 
11 square degrees in our opposition mini-survey). We followed or recovered 33 program NEAs 
(including 4 PHAs and 1 VI), making the INT-WFC one of the most productive EURONEAR 
facilities for recovery and survey work (NEAs and MBAs). In Section~\ref{stats} we use the 
INT findings for MBA and NEA statistics. 

\subsection{Blanco observing run}

The first EURONEAR run using a 4m telescope was awarded by the Chilean Time Allocation Committee
for 3--4 June 2011 (proposal number 0646, PI: E. Unda-Sanzana) to use the Victor Blanco 4m telescope 
at Cerro Tololo Inter-American Observatory (CTIO) in Chile. At the prime focus $F/2.9$ of the 
telescope we employed the MOSAIC II camera which consists of a $2\times4$ mosaic of CCDs $2K\times4K$ 
pixels each covering a total field of view of $36^\prime \times 36^\prime$ with a pixel scale of 
$0.27~^{\prime\prime}$/pix. To minimize the readout time and FTP data transfer, we used $2\times2$ 
binning (binned pixel size $0.54^{\prime\prime}$). We used an $R$ filter for the entire Blanco run. 
The filling factor of the MOSAIC II camera is 0.98.  We used normal sidereal tracking for all fields.

The time was dark, with only the first night clear and its first three hours affected by clouds, 
resulting in no NEA recovered in the first nine program NEA fields. The second night had complete 
cloud cover, one day before a big snow storm which hit Cerro Tololo. During the first 
night we recovered 12 program NEAs (including 1 VI and 9 PHAs) plus two other NEAs falling by 
chance in the observed fields (1 PHA). The typical seeing was $\sim1^{\prime\prime}$. 

For data reduction we used the USNO-B1 catalog and no field correction due to lack of time 
to fit THELI to work in binning mode with MOSAIC II. First, we reduced data allowing the high 
field distortion to be accommodated by a fit order 3 in Astrometrica. Then, the MPC advised us about 
some errors \citep{wil11} which we traced to the high field distortion of the prime focus camera, 
thus we restricted Astrometrica to a fit order 2 which allowed better astrometry. 

Besides NEAs (Table A1), we identified and measured all moving sources in all observed fields 
(except for four fields falling in the Milky Way), reporting all known and new objects visible 
up to $R\sim24$ mag. Using Blanco during one night, we observed in total 28 MOSAIC II fields 
covering in total 10 square degrees. In Section~\ref{stats} we use Blanco data to derive some 
MBA and NEA statistics.

\subsection{WHT observing runs}

A few observations to recover some NEAs were conducted during twilight or combined with some 
technical tasks during four service or discretionary (S/D) nights in 2011, using the William 
Herschel 4.2m Telescope (WHT) of the ING located at ORM. 
Additionally, during other two service nights in 2011 we observed for photometry the 
OSIRIX-Rex target (the VI (101955) 1999 RQ36) under the program SW2011a31 (PI: J. Licandro), 
for about 8 hours in total. In total using the WHT, we observed 11 program NEAs (including
4 PHAs and 1 VI). We used guiding only for the photometry runs of (101955), tracking at
normal sidereal rate for the other fields. 

At the Cassegrain $F/11$ focus we used the ACAM camera which consists of one CCD 
with $2K\times4K$ pixels covering a field of view of $8^\prime$ diameter with a pixel scale of 
$0.25~^{\prime\prime}$/pix. We used $R$ filters for all WHT runs. The weather was mostly good, 
with WHT ACAM typical seeing around $1^{\prime\prime}$. In Table A1 we list the program NEAs 
observed with WHT. 

For data reduction we used IRAF for typical flat and bias corrections and Astrometrica 
with a fit order 1-2 (sufficient for the Cassegrain instrument), using mostly USNO-B1, UCAC3 
or UCAC4 depending on the star density in the observed small ACAM field. 

\subsection{TLS observing runs}

T\"uringer Landessternwarte Observatory (TLS) located in Tautenburg, Germany, joined the EURONEAR 
network in 2009, starting regular NEA observations in 2011 using the Alfred Jensch 2m telescope 
(1.34m Schmidt corrector plate). At the Schmidt $F/4$ focus of the TLS telescope is a SITe CCD 
$2k\times2k$ camera, covering a large field of view $42^\prime \times 42^\prime$ with a pixel 
scale of $1.23~^{\prime\prime}$/pix. 

During 39 nights in September 2011, March, and May 2012, a total of 38 NEAs, mostly very recently 
discovered objects, were observed (Table A1). The median seeing in Tautenburg is $\sim 2^{\prime\prime}$ 
and the $R$ filter was used for all runs. The TLS pipeline which includes run planning and data 
reduction is complete, and few runs are planned in the near future. 

Astrometry and field correction are resolved using $Astrometry.net$ \citep{lan10} 
with the GSC 2.3 catalog. The exposure time used at TLS was a standard 180 sec for all NEAs. 
The observing technique and data reduction method are a novelty in
EURONEAR, so we briefly present this here.
Depending on the magnitude and proper motion of the object, a few frames are taken with 
same exposure time in normal sidereal auto-guiding, thus the asteroid becomes visible as a trail. 
After basic correction of bias and flat field, field distortion and astrometry are resolved with 
$astrometry.net$, using second order SIP polynomials to account for field distortion. The median 
co-added image is created by registering all images on the center of the NEA trail given the shifts 
calculated from the ephemeris. Finally, the NEA trail will be deconvolved using a PSF artificial 
trail (length and direction from motion, width from stellar PSF) to model the centrally peaked 
elongated brightness distribution of the object. Finally, the co-added trail is fitted by an 
elliptical Gaussian to come up with object coordinates and errors. 

\begin{figure}
\centering
\fbox{\includegraphics[angle=0,width=13.5cm]{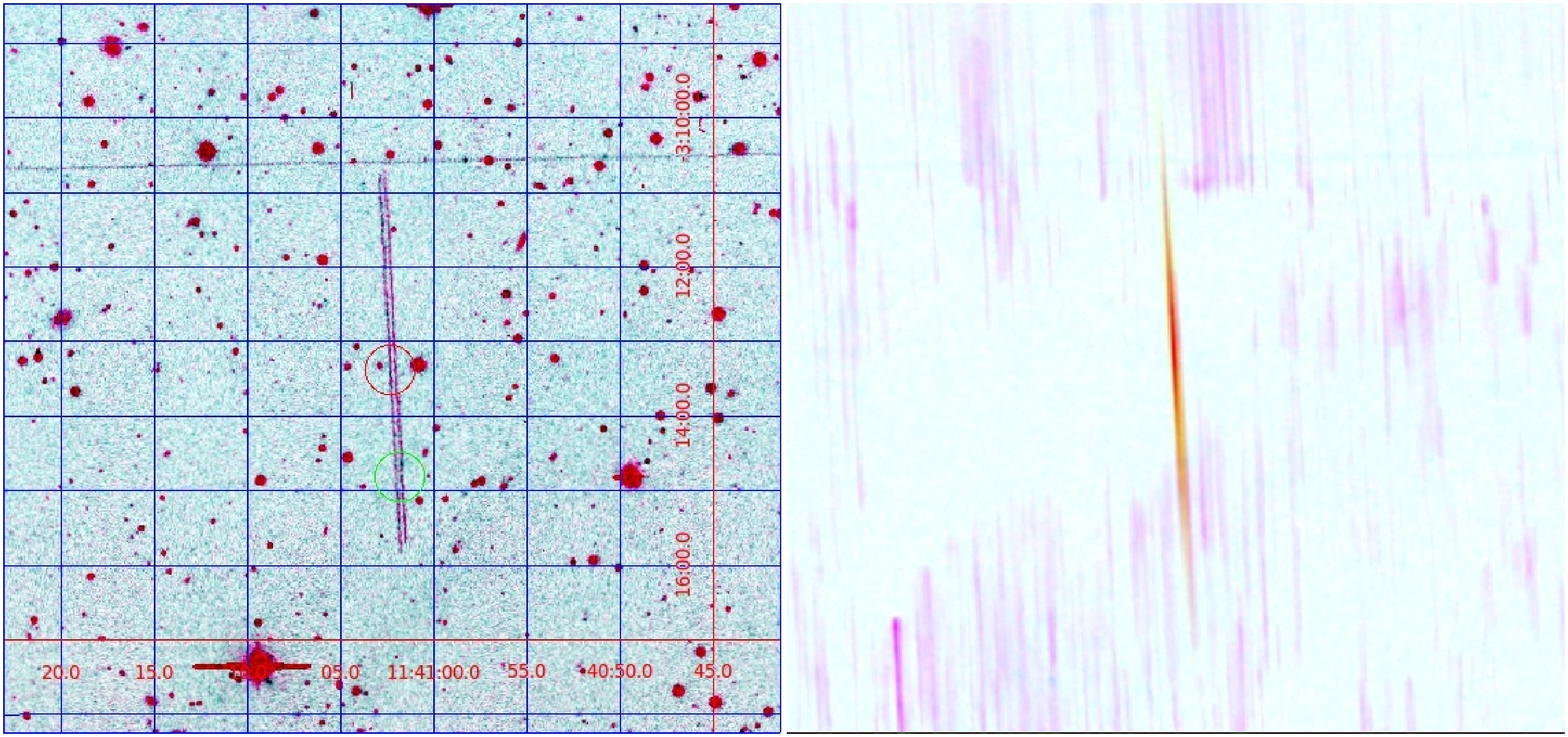}}
\begin{center}
\caption{The very fast NEA 2012 FP35 observed at TLS using the trail deconvolution reduction 
method. On the left side we present the sky image superposed on the predicted two trails 
of the object. On the right side we present the recovered (final) image obtained by co-addition 
of the two observed images registered on the NEA trail based on the ephemeris. }
\label{fig1}
\end{center}
\end{figure}

Figure~\ref{fig1} shows the case of the extremely newly discovered (one day) fast NEA 2012~FP35 
moving with $\mu=100.5^{\prime\prime}$/min. The left image is the result of adding two registered 
frames, resulting in two observed trails which do not exactly superimpose because of the large
uncertainty in the orbit (e.g. position and proper motion). The predicted position is indicated 
by the green circle while the observed one (the result of the fit) is marked by the red circle. 
The right image represents the recovered deconvolved image where the stars become elongated. 

\subsection{OHP observing runs}

Two observing runs (10 nights in total) were awarded in 20--24 April 2010 and 15--19 November 2010 
by the Local Time Allocation Committee using the 1.2m telescope (T120) at Haute de Provence Observatory 
(OHP) in France. 

The Netwonian $F/6$ T120 telescope hosted the TK 1024 AB camera equipped with a $1k\times1k$ 
CCD covering a $11.7^\prime \times 11.7^\prime$ field of view with a pixel scale of $0.68~^{\prime\prime}$/pix. 
The median seeing at OHP was about $2.5^{\prime\prime}$. The $R$ filter was used for both runs 
and normal sidereal tracking. For data reduction we used Astrometrica with fit order 1 and 
USNO-B1 catalog to provide enough stars for the relatively small and shallow observed fields. 

Past (pre-2010) OHP observations were reported by \cite{bir10a}. In total during the 2010 runs we 
observed 36 program NEAs (including 8 PHAs; Table A1).

\subsection{Pic du Midi observing runs}

Two observing runs (10 nights in total) were awarded in 1--4 March 2011 and 17-24 November 2011 
by the {\it Station de Planetologie des Pyrenees} managed by {\it Observatoire de Paris} and 
{\it Observatoire Midi Pyrenees} at Pic du Midi Observatory, France. For both runs the T1m 1.05m 
telescope was used. 

For the first run, at the Cassegrain $F/12.1$ reduced focus we employed the iKon-L Andor camera with 
a $2k\times2k$ E2V chip with pixel scale $0.22~^{\prime\prime}$/pix. To reduce the noise 
we used $2\times2$ binning, so a pixel size of $0.44^{\prime\prime}$ with a field of view of 
$7.5^\prime \times 7.5^\prime$. 
For the second run at the Cassegrain $F/7.6$ reduced focus we employed the Atik 383L+ camera with 
an $3326\times2504$ CCD Kodak KAF-8300 chip with pixel scale $0.14~^{\prime\prime}$/pix. We used 
$3\times3$ binning, so a pixel size of $0.42^{\prime\prime}$ with a field of view of 
$7.8^\prime \times 5.8^\prime$. 
For both runs we used a clear filter from Astronomik equivalent to $B+V+R$ with bandpass from 390 nm 
to 680 nm, tracking at half the proper motion of observed NEAs. 

Typical seeing at Pic du Midi was about $1.5^{\prime\prime}$. The weather was good and the 
Moon was mostly dark. 
For the data reduction we used Astrometrica with fit order 1, enough for the Cassegrain field, 
and USNO-B1 and NOMAD catalogs, depending on available star density in the fields. 

Past EURONEAR Pic du Midi observations were reported in \cite{vad08} and \cite{bir10a}. During 
the two Pic du Midi 2011 runs, 34 NEAs were observed (including 5 PHAs and 1 VI; Table A1).

\subsection{Bonn/AIfA observing runs}

The Argelander Institute for Astronomy (AIfA) owns a 0.5m telescope hosting at its Cassegrain $F/9$ 
focus an SBIG-STL 6303E CCD camera ($3072\times2048$ pixels, scale $0.45^{\prime\prime}$/pixel).
The typical seeing there is about $3^{\prime\prime}$ and the light pollution is quite high (naked-eye 
mag 4 at zenith) allowing mag 19 to be reached only after stacking a few short exposed images (using 
Astrometrica's ``track and stack'' capability). We observed with $2\times2$ binning using $V$ and 
$R$ filters, tracking all fields at normal sidereal rate. We used Astrometrica with UCAC3 and 
USNO-B2 catalogs and no field correction in the Cassegrain field. 

The MPC code of the Bonn/AIfA observatory C60 was obtained by A. Tudorica (MSc student there) who 
joined the Bonn/AIfA node to the EURONEAR network in 2011. During 25 nights between September 2011 
and May 2012, we observed 60 NEAs (of which 15 PHAs) and reported 830 positions.

\subsection{Galati observing runs}

The public Astronomical Observatory of the Natural Sciences Museum Complex in the city of Galati, 
Romania, was founded thanks to the efforts of O. Tercu, the actual observatory coordinator, based on 
funding from the county of Galati ({\it Consiliul Judetean Galati}), plus other funding from the 
E.U. and some local sponsorship. The observatory joined the EURONEAR network in 2010 and
obtained the MPC code C73 in 2011. Supported by a small but enthusiatic team of amateurs
in the local astronomy club ``Calin Popovici'', Galati observatory includes NEAs as one of their 
main priorities for public outreach and education. 

Among other smaller telescopes and instruments, the main telescope of Galati observatory is the 
ASA 0.4m Ritchey-Chretien f/8 telescope endowed with an SBIG STL-6303E CCD with $3072\times2048$ 
pixels covering a relatively large field of view $29.8^\prime \times 19.9^\prime$ with pixel scale 
$0.58^{\prime\prime}$/pix. 

All NEA observations of Galati Observatory were performed using 
$2\times2$ binning (binned pixel size $1.16^{\prime\prime}$/pix), due to the typical seeing 
of about $3^{\prime\prime}$. For all runs we used normal tracking at sidereal rate. 
For data reduction we used Astrometrica with UCAC3 catalog and a fit order 4 to correct the 
relatively distorted large field. 

During 55 nights between Sep 2011 and May 2012 three people observed 223 NEAs (of which 
40 PHAs and 4 VIs) and reported 2367 positions, becoming the most prolific EURONEAR site in 
terms of number of observed objects and NEA public outreach. Despite some city light 
pollution, the 40-cm telescope manages to reach limiting magnitude $V\sim19$. 

\subsection{Urseanu observing runs}

The public ``Admiral Vasile Urseanu Observatory'' in Bucharest, Romania, was founded in 1910 
by the navy admiral Vasile Urseanu who was also an amateur astronomer. The actual public meeting 
place of many visitors and host of the Bucharest Astroclub, the observatory is owned by the 
Bucharest Museum of History. Thanks to the efforts of A. Sonka, in 2006 the observatory obtained 
its MPC code A92, becoming one of the first EURONEAR nodes.

The very small 0.3m telescope (Meade LX200R) of the Urseanu Observatory is actually the 
smallest used by our network, being extensively used for public outreach there. At the Cassegrain 
F/6.3 field resides an inexpensive QHY6 CCD camera. The pixel scale is $2.30^{\prime\prime}$/pix 
and the field is $21.6^\prime \times 16.9^\prime$. Typical seeing in the very light polluted 
capital Bucharest using this equipment is about $3^{\prime\prime}$ and the limiting magnitude 
is $V\sim16$ ($V\sim15$ for fast moving objects). For all runs we used sidereal tracking. 

Past NEA observations from Urseanu Observatory were reported in \cite{vad08} and \cite{bir10a}. 
From November 2008 until May 2012 a total of 28 NEAs (including 9 PHAs and 1 VI) were observed 
and reported from this observatory. 

%% ==================================================================================================

\section{Results}
\label{results}

\subsection{Program NEAs}
\label{NEA}

\begin{figure}[p]
\centering
\includegraphics[angle=0,width=6.7cm]{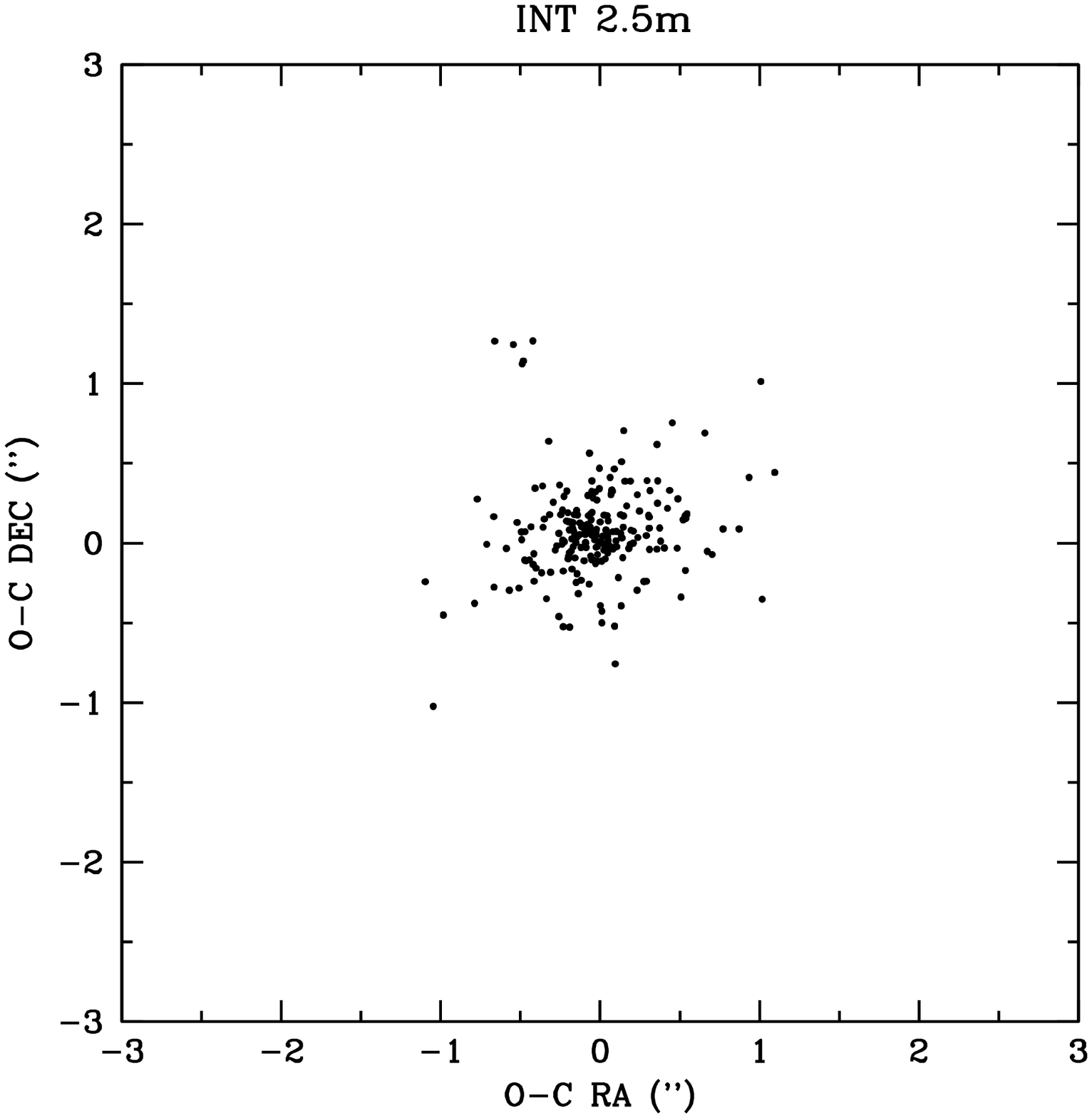}
\includegraphics[angle=0,width=6.7cm]{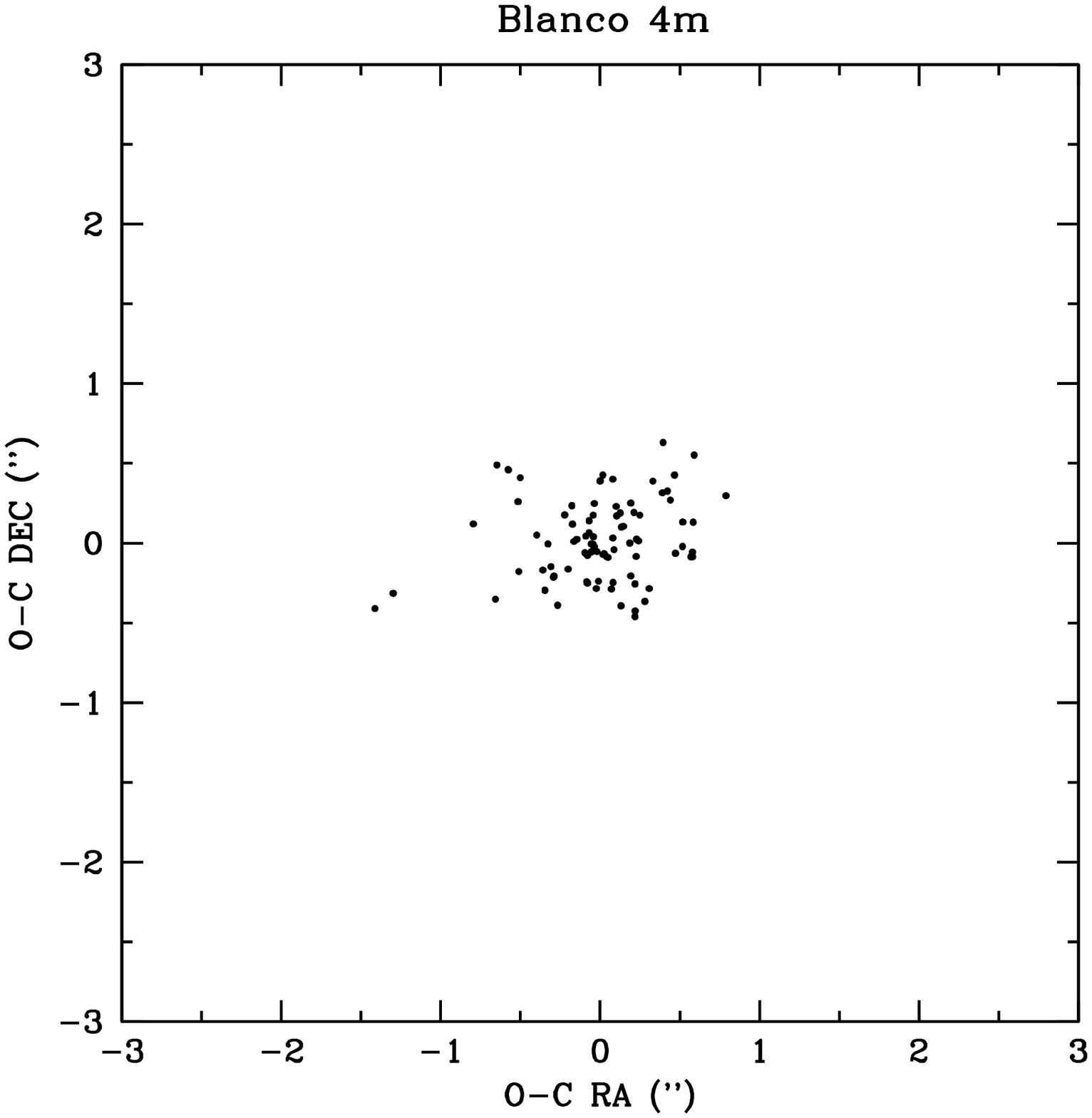} \\
\includegraphics[angle=0,width=6.7cm]{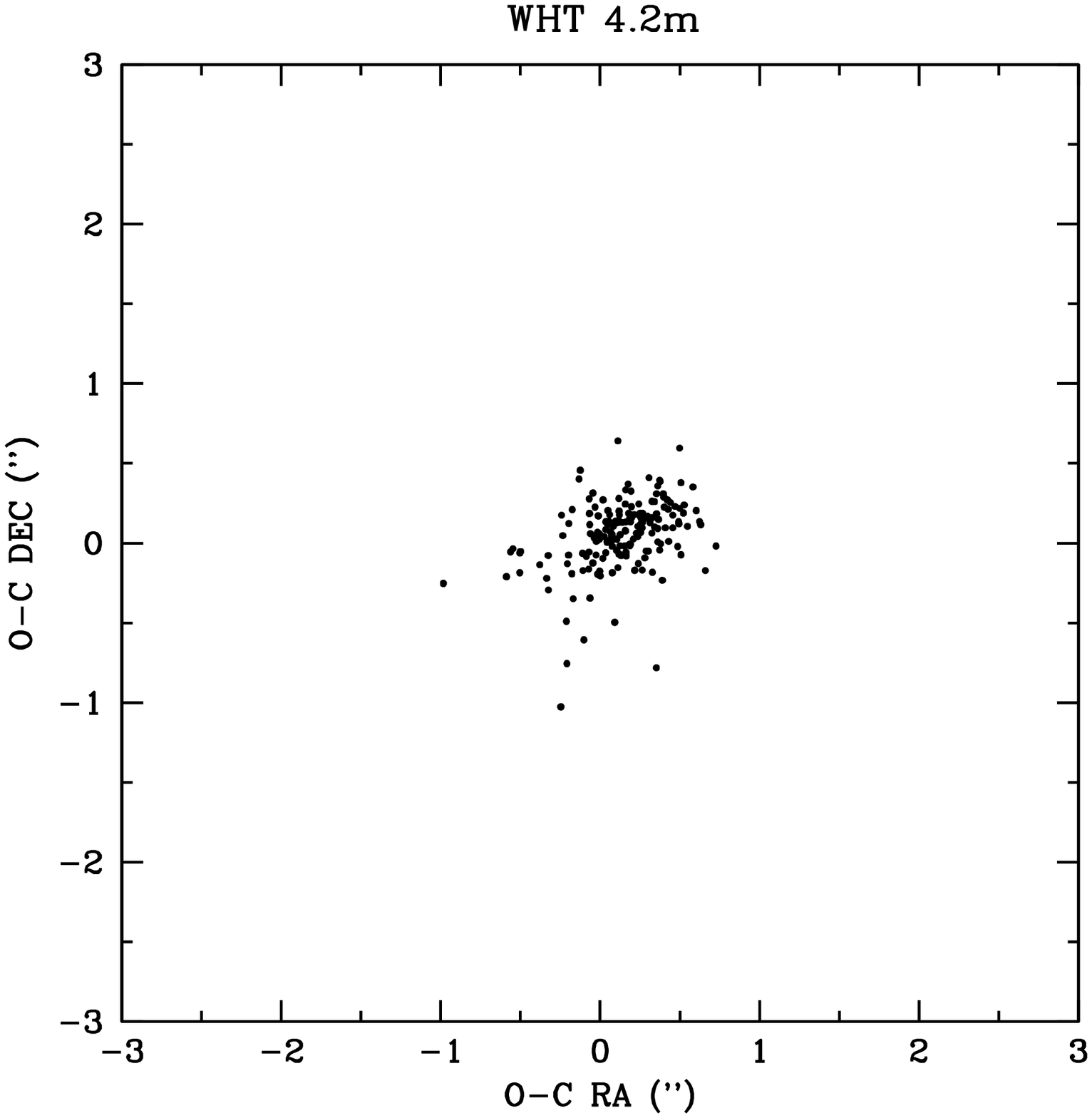} 
\includegraphics[angle=0,width=6.7cm]{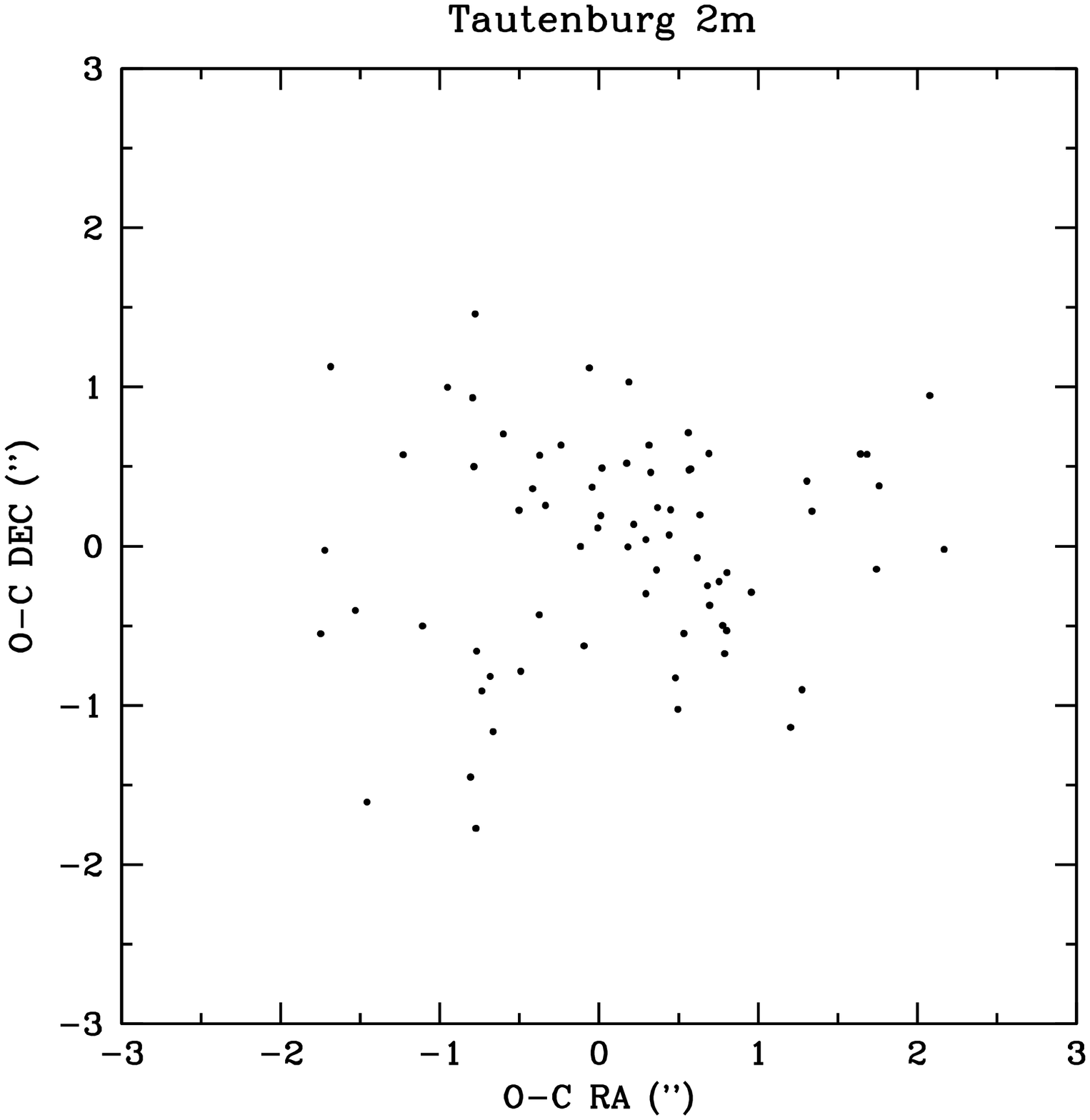} \\
\includegraphics[angle=0,width=6.7cm]{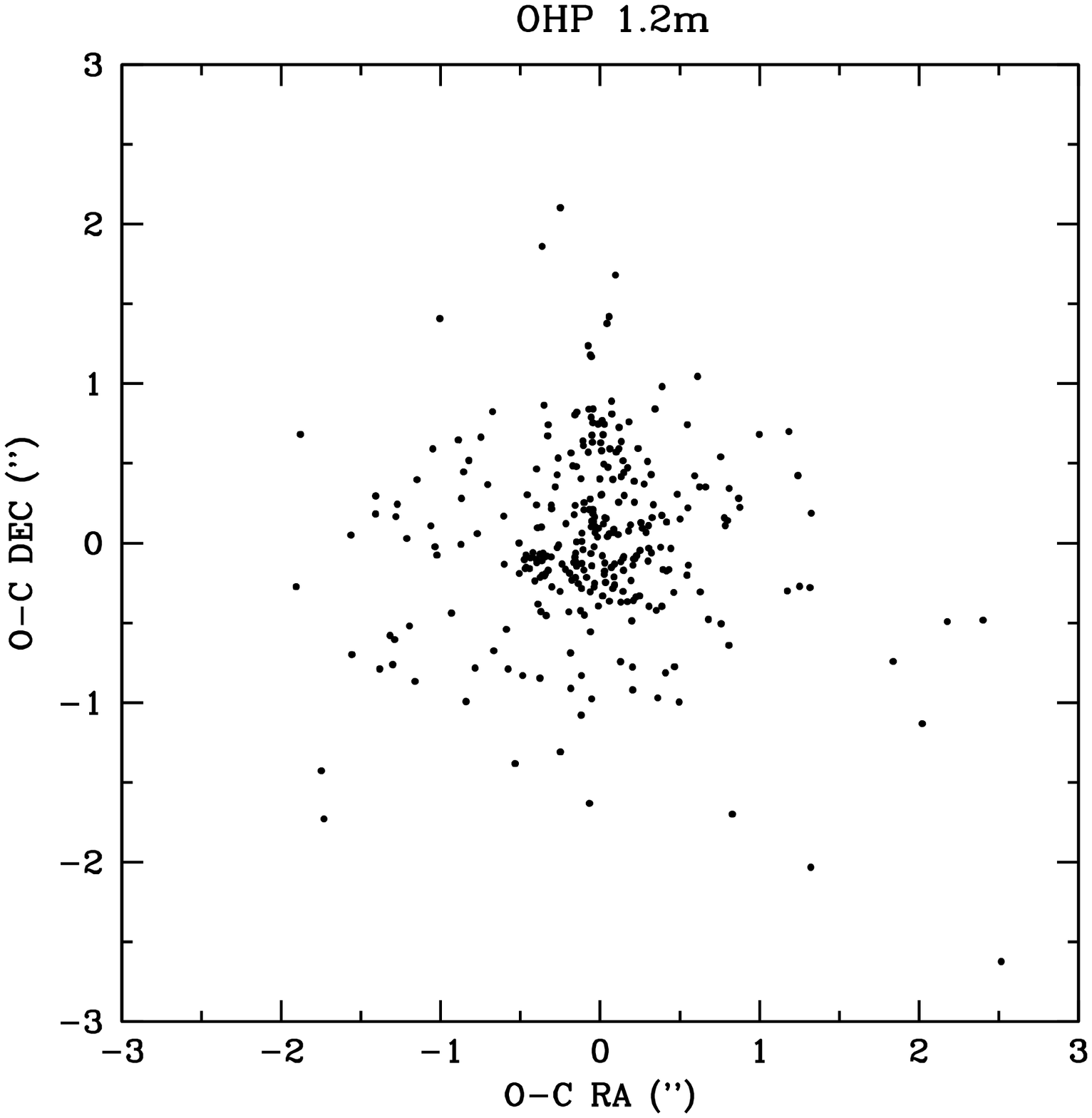}
\includegraphics[angle=0,width=6.7cm]{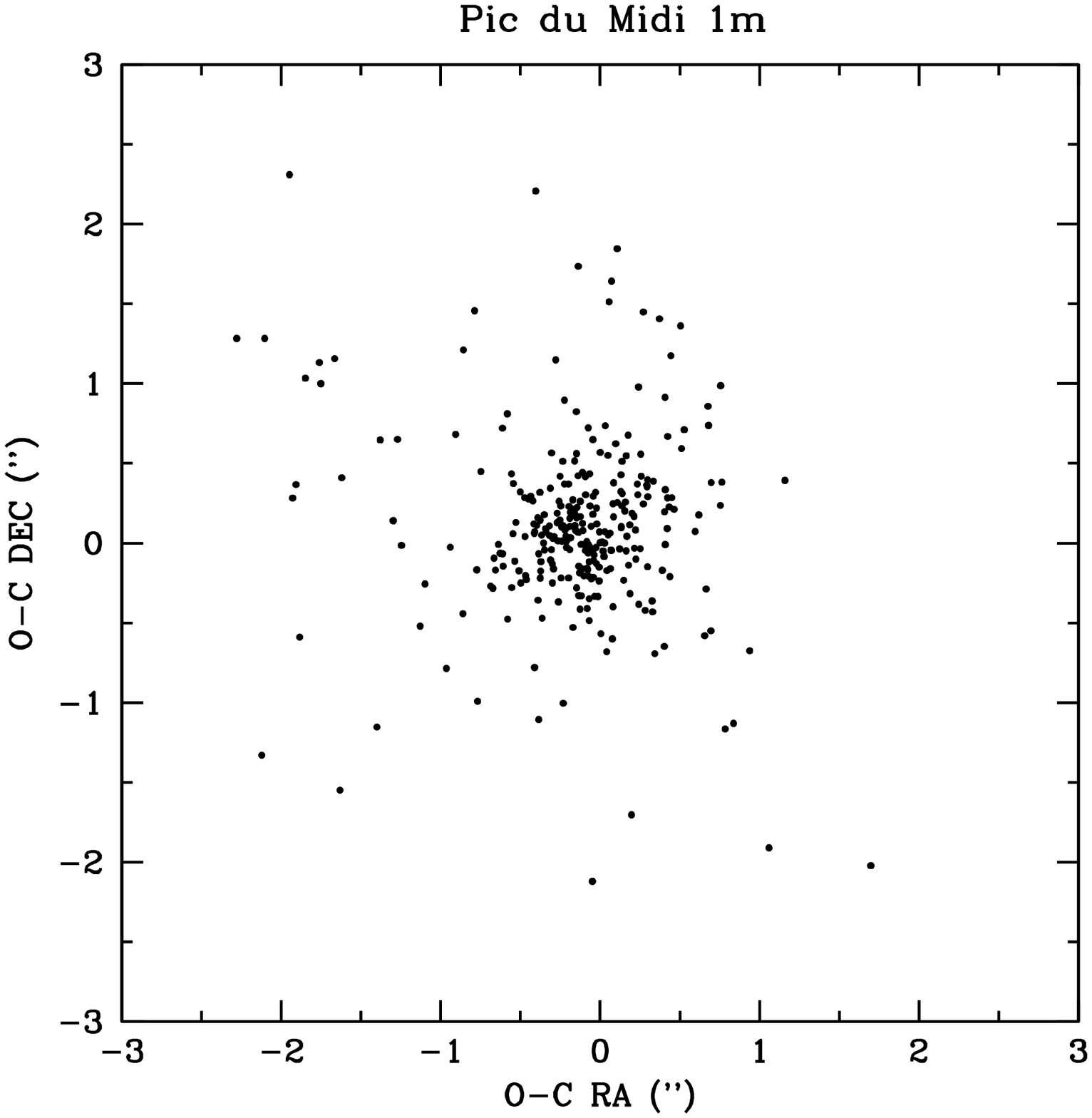} \\
\begin{center}
\caption{O--C (observed minus calculated) residuals for program NEAs observed using the telescopes presented in this paper. } 
\label{fig2}
\end{center}
\end{figure}

A total of 477 program NEAs observed within the EURONEAR network between 2009 
and May 2012 are reported in this paper. 
Of these, 166 were observed with six professional 1-4m telescopes and 311 with 
three smaller (0.3-0.5m) educational and public outreach telescopes. 
In the Appendix Table A1 we give the observing logs containing these objects observed with the 
above nine telescopes. In Figure~\ref{fig2} we plot the O--C (observed minus calculated) 
residuals in $\alpha$ and $\delta$ for the program NEAs observed with each telescope. 

\begin{figure}[p]
\centering
\includegraphics[angle=0,width=6.7cm]{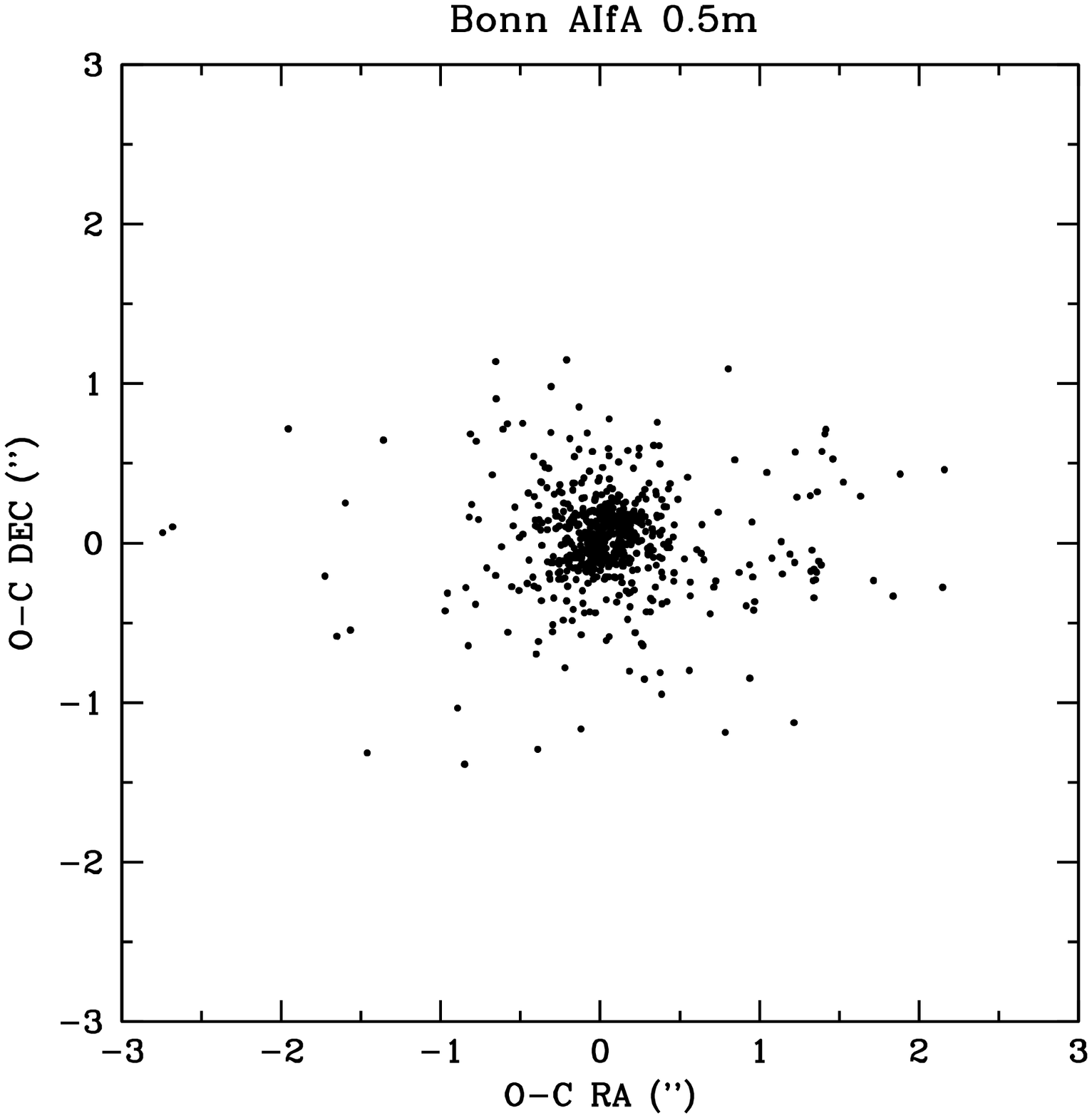} 
\includegraphics[angle=0,width=6.7cm]{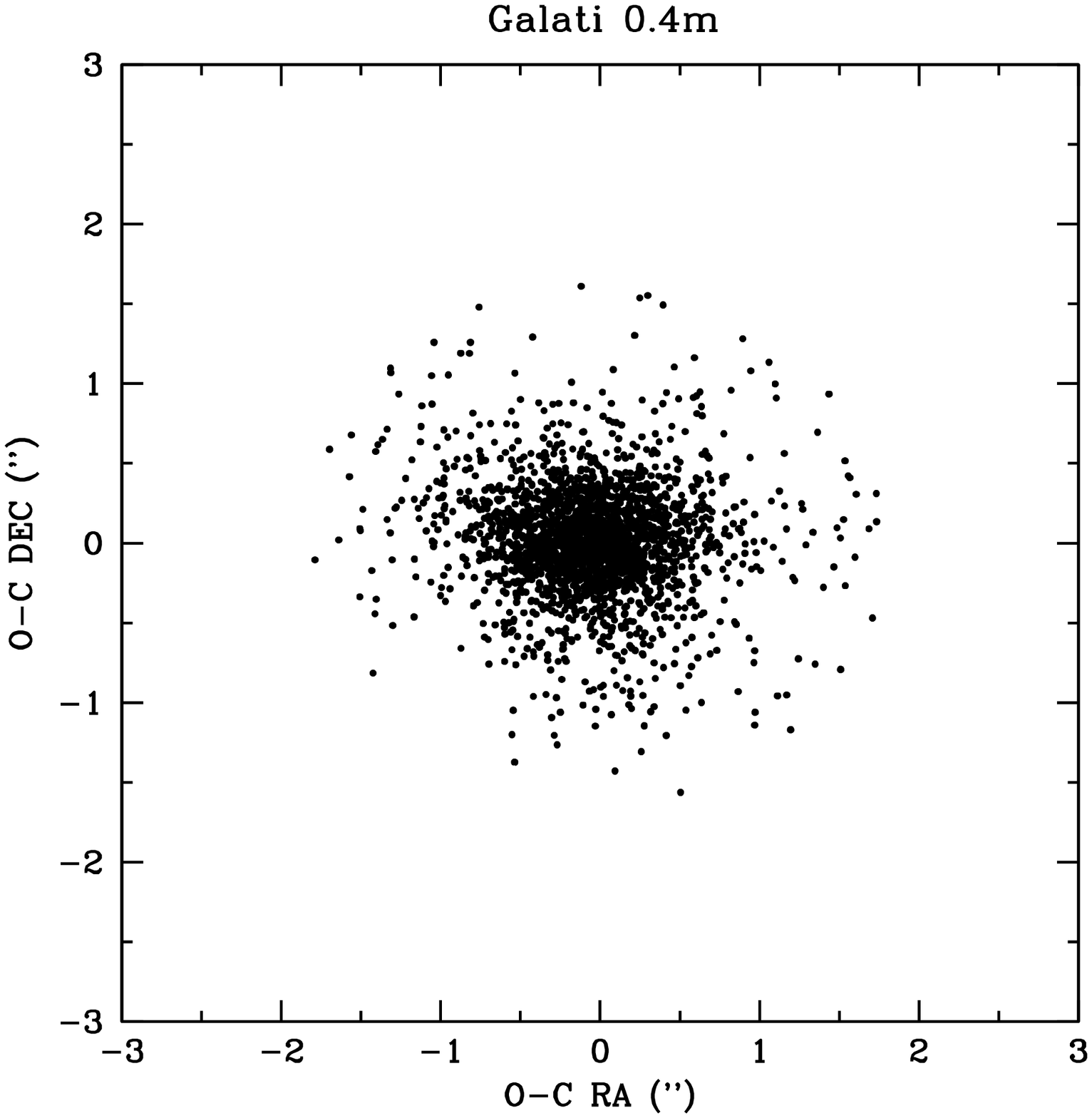} \\
\includegraphics[angle=0,width=6.7cm]{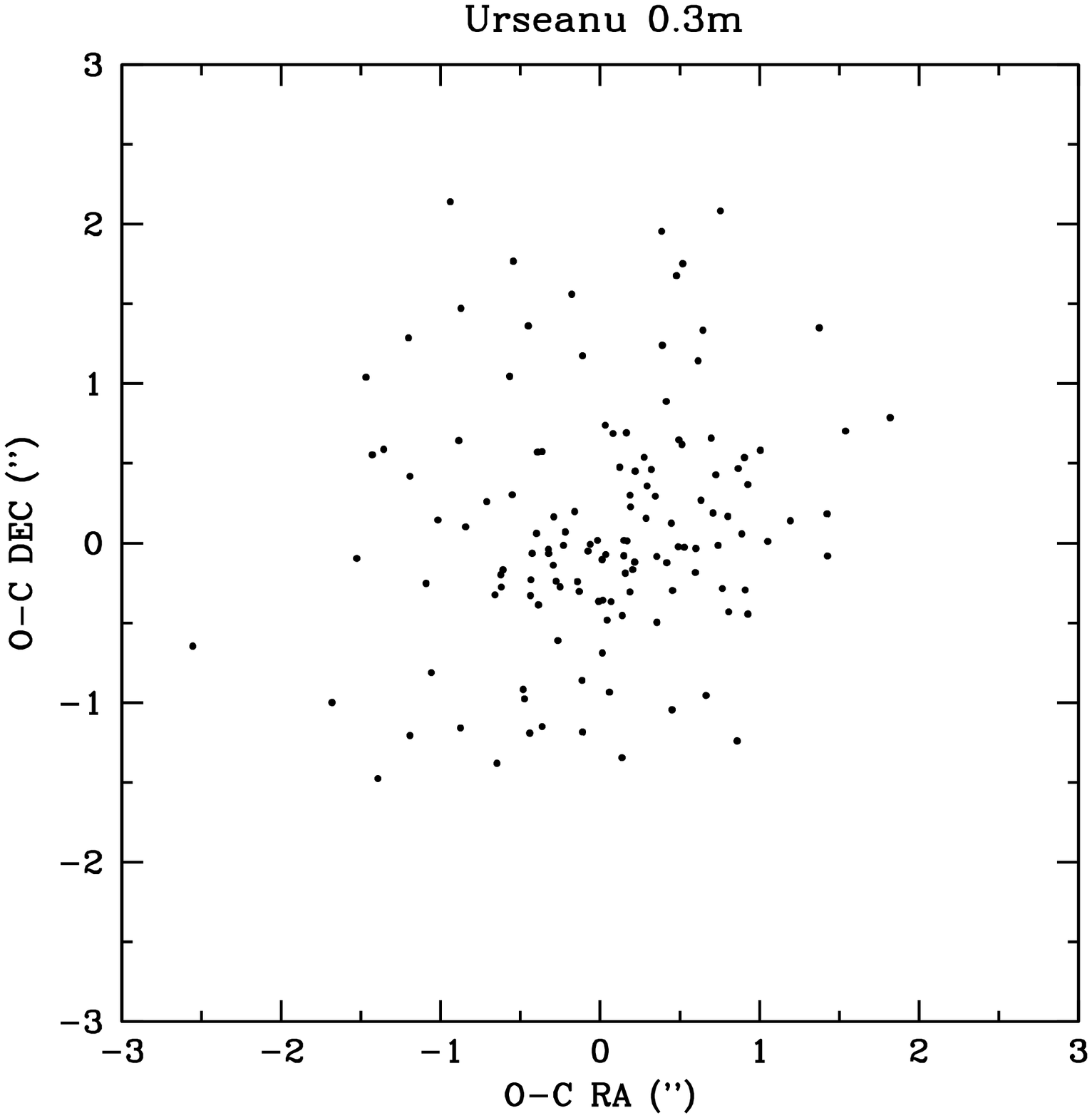}
\begin{center}
Figure 2 (continued)
\end{center}
\end{figure}

The root mean square of the O--C residuals for our MPC published NEA datasets for each telescope are, in order: 
WHT $0.44^{\prime\prime}$, INT $0.46^{\prime\prime}$, Blanco $0.46^{\prime\prime}$, Bonn $0.56^{\prime\prime}$, 
Galati $0.57^{\prime\prime}$, Pic $0.81^{\prime\prime}$, OHP $0.85^{\prime\prime}$, Urseanu $1.05^{\prime\prime}$ 
and TLS $1.14^{\prime\prime}$. 

Overall, small amateur-educational telescopes Bonn and Galati are performing very good astrometry, 
only slightly inferior by $0.1^{\prime\prime}$ than their much larger 2-4m ``colleagues'' INT, Blanco 
and WHT which also benefit from much better weather conditions. Owing to less accurate catalogs used 
(USNO-B1 versus UCAC3) and possibly due to target selection (newly discovered versus new opposition 
objects) we observe that O-Cs of our two 1m facilities OHP and Pic trail behind Bonn and Galati by about 
$0.3^{\prime\prime}$. Thus, the loss in astrometric precision is absolutely non linear with the 
shortening of aperture, the most important factors favouring the contribution from small telescopes being 
better astrometric catalogs used (especially in the bright regime more accessible to small telescopes), 
a good SNR for both target and reference stars and the absence of complicated field distortion, all 
in line with recommendations of the IAU Working Group ``Astrometry by Small Ground Based Telescopes''. 

At least two inhomogeneities can be spotted in the O-C plots in Figure~\ref{fig2}. First, the small 
vertical branch in the OHP plot is due to few OHP objects moving much faster in $\delta$ than in $\alpha$, 
resulting in larger measurement errors and O-Cs in $\delta$. Second, the WHT O-C centroid seems slightly 
displaced (by about $0.2^{\prime\prime}$) to the upper-right from the
origin, which is explained by the known systematics of the mostly used
USNO-B1 catalog due to the small ACAM field (a fact observed also by us
previously \citep{bir10a}).

\subsubsection{MPC/MPEC publications}

Adding the 477 observed NEAs to our previous work presented in \cite{bir10a} and \cite{vad11}, 
we report a total of 739 NEAs observed within the EURONEAR network during the last six years. 
The reduced data presented in this paper generated 98 MPC publications,
comprising 69 Minor Planet Circulars (MPC) and 29 Minor Planet Electronic Circulars (MPEC).

The Blanco run resulted in published data in the following eight MPC/MPEC publications: 
74496, 75198, 75623, 75939, 76865, 79528 (Buie et al: 2011a, 2011b, 2011c, 2011d, 2011e, 2012), 
76441 and 77265 (Elst et al: 2011a, 2011b). 

The INT runs resulted in published data in the following 11 MPC/MPEC publications: 
76443, 78047, 78437, 78894, 79221, 79530 (Fitzsimmons et al: 2011a, 2012a, 2012b, 2012c, 
2012d, 2012e), 75625 (Holman et al, 2011), 79787 and 2012-D82 (Vaduvescu et al: 2012a, 2012b), 
2012-E19 (Gajdos et al, 2012) and 2012-D102 (Bressi et al, 2012). 

The WHT observations resulted in published data in the following eight MPC/MPEC publications:
74148, 74893, 75207, 76443, 77266, 78047 (Fitzsimmons et al: 2011a, 2011b, 2011c, 2011d, 
2011e, 2012a), 75625 (Holman et al, 2011) and 75940 (Balam et al, 2011). 

The OHP runs resulted in published data in the following six MPC/MPEC publications:
69732, 72456 (Arlot et al: 2010a, 2010b), 70198, 2010-W11, 2010-W12, 2010-W13 
(Birlan et al: 2010b, 2010c, 2010d, 2010e). 

The Pic du Midi runs resulted in published data in the following 17 MPC publications:
74036, 77173 (Cavadore et al: 2011a, 2011b), 2011-E11, 2011-E12, 2011-E13, 2011-W29, 2011-W33
(Birlan et al: 2011a, 2011b, 2011c, 2011d, 2011e), 2011-E14 (Lehman et al, 2011), 2011-E19
(Apitzsch et al, 2011), 2011-W25 (McMillan et al, 2011), 2011-W12, 2011-W22, 2011-W27, 
2011-W28, 2011-W44, 2011-W45 and 2011-W52 (Buzzi et al: 2011a, 2011b, 2011c, 2011d, 
2011e, 2011f, 2011g). 

Tautenburg runs resulted in published data in the following six MPC/MPEC publications:
MPC 79140, 79473 (Borngen and Stecklum: 2012a, 2012b), 79746, 79971 (Stecklum, 2012a, 2012b), 
2012-G45 and 2012-K07 (Stecklum et al, 2012a, 2012b). 

Bonn AIfA runs resulted in published data in the following nine MPC/MPEC publications:
MPC 74509, 79224, 79789, 80000 (Tudorica, 2011, 2012a, 2012b, 2012c), 76451 (Tudorica and Toma, 2011), 
76874, 78902 (Tudorica and Badescu: 2011, 2012), 77273 and 79533 (Tudorica et al, 2012d, 2012e). 

The Galati runs resulted in published data in the following 17 MPC/MPEC publications: 
MPC 75947, 76451, 76874, 77273, 77705, 78053, 78445, 78902, 79225, 79533, 80000
(Tercu and Dumitriu: 2011a, 2011b, 2011c, 2011d, 2012a, 2012b, 2012c, 2012d, 2012e, 2012f, 2012g), 
2012-B17 (Bacci et al, 2012), 2012-G45 (Stecklum et al, 2012), 2012-H42 (Eglitis et al, 2012), 
2012-H90 (Jaeger et al, 2012), 2012-J01 (Nishiyama et al, 2012) and 2012-J34 (Christensen et al, 2012). 

Finally, Urseanu runs resulted in published data in the following 16 MPC/MPEC publications: 
MPC 64104, 65048, 65638, 65928, 67139, 67404, 67677, 69817, 71094, 72053, 75208, 75447,
77270, 77702, 78050 and 79998 (Sonka: 2008, 2009a, 2009b, 2009c, 2009d, 2009e, 2009f, 2010a, 2010b, 
2010c, 2011a, 2011b, 2011c, 2012a, 2012b, 2012c).

\begin{figure}
\centering
\includegraphics[angle=0,width=7cm]{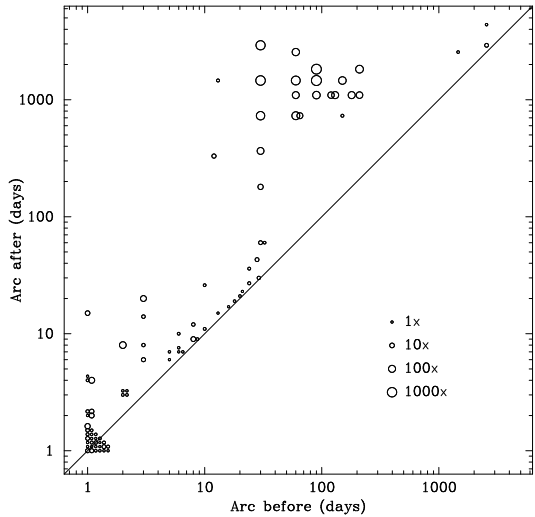}
\caption{Orbital improvement as a function of observational timespan.
Distance above diagonal line corresponds to factor by which observed arc has increased.
Size of each circle shows factor by which uncertainty in $a$ has improved as a result of 
increase in observed arc. Points near (1 day, 1 day) have been slightly separated for clarity.} 
\label{sig_a}
\end{figure}

\subsubsection{Orbital improvement}

We focus next on the NEAs recovered at a new opposition (29 objects marked by $\star$ in Appendix 
Tables A1 and A2) and also on the NEAs followed-up soon after discovery - new objects whose arcs were 
either observed within the first day or prolonged by at least one day during the first month (82 objects 
marked by $\bullet$ in Tables A1 and A2), studying the orbits of these 111 objects. 

We use the software FIND\_ORB \citep{gra12} to assess the orbital improvement, by comparing the 
orbital elements fitted with observational data available before our runs (first line in
Table A2) with those obtained adding our observations (second line in Table A2). 
We list in the table the semimajor axis $a$, eccentricity $e$, inclination $i$, 
longitude of ascending node $\Omega$, argument of pericenter $\omega$, mean anomaly $M$ and 
minimum orbital intersection distance MOID. 

Despite the importance of the recovery of NEAs at a new opposition which
lengthens the observed arc by a few years, orbital elements of the NEAs
recovered at a new opposition appear to be improved only slightly in
absolute terms: standard deviations of the changes to elements 
(based on 25 pairs of orbits for objects marked $\star$ in Table A2) are
$a\sim10^{-4}$ AU, $e\sim5\times10^{-5}$, $i\sim10^{-3}\deg$, $\Omega\sim10^{-3}\deg$, 
$\omega\sim3\times10^{-3}\deg$, $M\sim5\times10^{-2}\deg$, MOID $\sim 3\times10^{-5}$ AU and 
$\sigma\sim0.02^{\prime\prime}$ (RMS of the fit).

Orbits of NEAs followed soon after discovery,
as they are less well constrained a priori, naturally undergo larger
changes to the elements following our observations, with 
standard deviations of $a\sim10^{-1}$ AU, $e\sim10^{-2}$, $i\sim1\deg$, $\Omega\sim1\deg$, 
$\omega\sim1\deg$ (all about $10^3$ times more), $M\sim7\deg$, MOID=$0.0151$ AU and 
$\sigma\sim0.06^{\prime\prime}$, based on 73 pairs of orbits representing objects observed with 
all telescopes marked $\bullet$ in Table A2.

FIND\_ORB additionally provides the option to calculate
uncertainties in the orbits it determines, enabling us to assess the
orbital quality both before and after our observations (as opposed to
simply the change in the best estimate orbit due to our observations).
We therefore calculated 1-sigma $a$, $e$ and $i$ uncertainties
$\sigma_a$, $\sigma_e$ and $\sigma_i$ for the Table A2 objects and can
define orbital improvement as the ratio in the $\sigma_a$, $\sigma_e$ or
$\sigma_i$ values before and after our EURONEAR observations. Figure
\ref{sig_a}, and similar plots for $e$ and $i$ not shown, suggest the
orbital improvement is closely related to the increase in time interval
spanned by the observations. This can be quantified by fitting log of
orbital improvement to log of proportional increase in timespan. The
formal linear regression results were:\\

\noindent
\(
\log (\sigma_a \; \mathrm{ratio}) = (1.35\pm0.22) \log T + (0.20\pm0.15) \\
\log (\sigma_e \; \mathrm{ratio}) = (0.81\pm0.19) \log T + (0.26\pm0.13) \\
\log (\sigma_i \; \mathrm{ratio}) = (0.57\pm0.15) \log T + (0.21\pm0.10) \\
\)

where T is the ratio of observed timespan after to before, and $\pm$
refers to 95\% confidence intervals. Considering separately the objects
observed in the days to weeks after discovery, the $\sigma_a$,
$\sigma_e$ and $\sigma_i$ regression slopes were all close to 1.0. For
the objects recovered at a new opposition, the slope was 1.0 for
$\sigma_a$, and 0.6 for $\sigma_e$ and $\sigma_i$ (but 1.0 marginally
permitted at 95\% confidence for both). A value of 1.0 would imply that
the proportional orbital improvement equals the proportional increase in
observed timespan. In practice, the number and distribution of
observations within the timespan, and the accuracy of the observations,
also affect the orbital quality (cf.\ \cite{mbw94}).

Based on the analyzed datasets, we conclude that the most important orbital contribution is the 
rapid follow-up of newly discovered NEAs (few days to one month after discovery). Another contribution 
is the recovery of NEAs at a new opposition which enlarges orbital arcs by a few months or years. 
Obviously, VIs and PHAs have priority over regular NEAs. 
Most important are fainter objects ($V>21$) which could remain invisible to the available 1-2m 
surveys and could result in insecure orbits or lost objects not being observed for the next few years.

\subsection{Known MBAs}
\label{MBA}

Many known MBAs were serendipitously encountered, identified, measured and reported in the 
large field images taken by Blanco MOSAIC II and INT WFC cameras. In total, 1,699 positions 
corresponding to 288 known MBAs were reported during our Blanco Mosaic II run, while 3,465 
positions of 580 known MBAs were reported from the observed INT-WFC fields. 

\begin{figure*}
  \centerline{
    \mbox{\includegraphics[width=7cm]{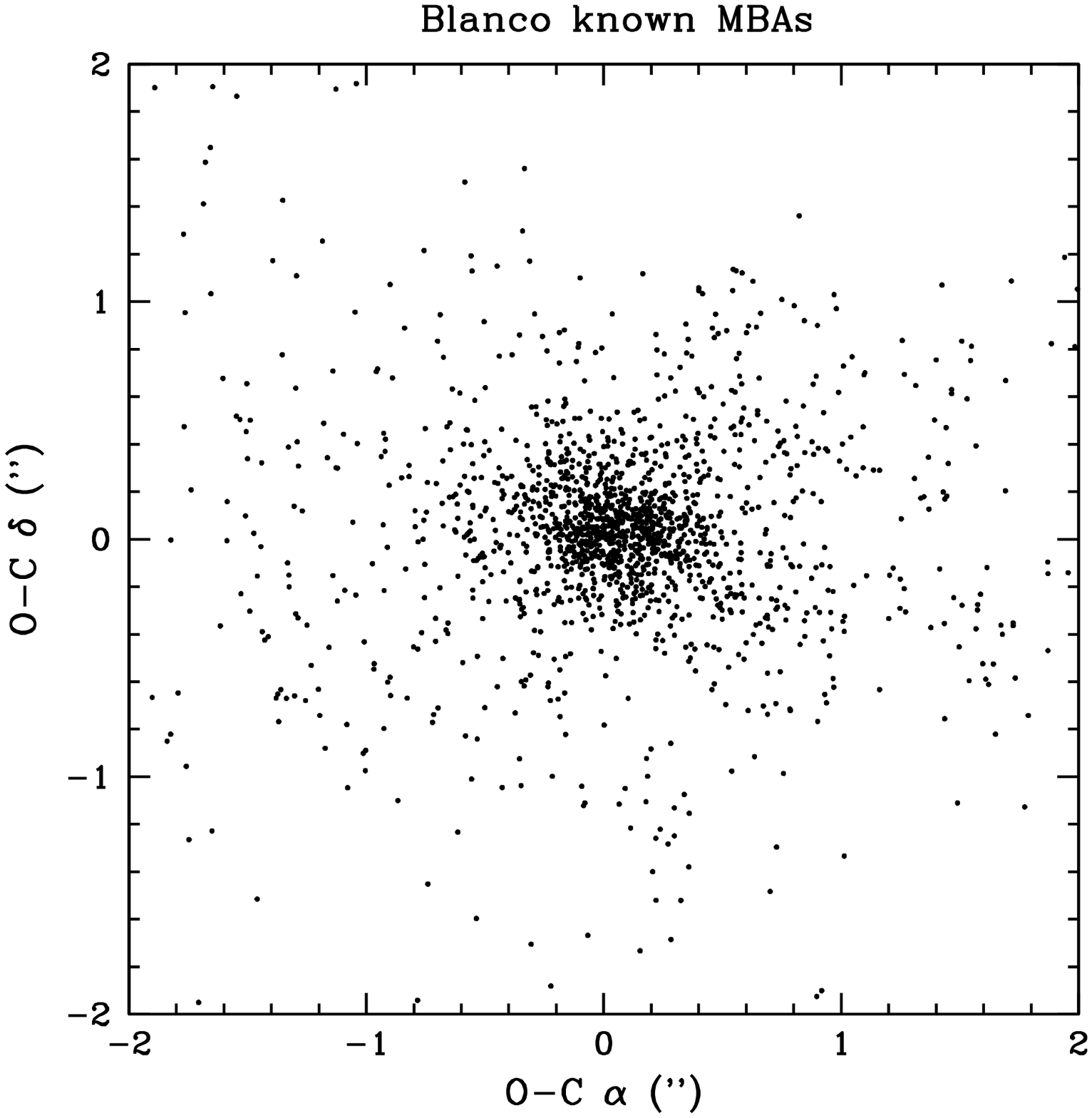}}
    \mbox{\includegraphics[width=7cm]{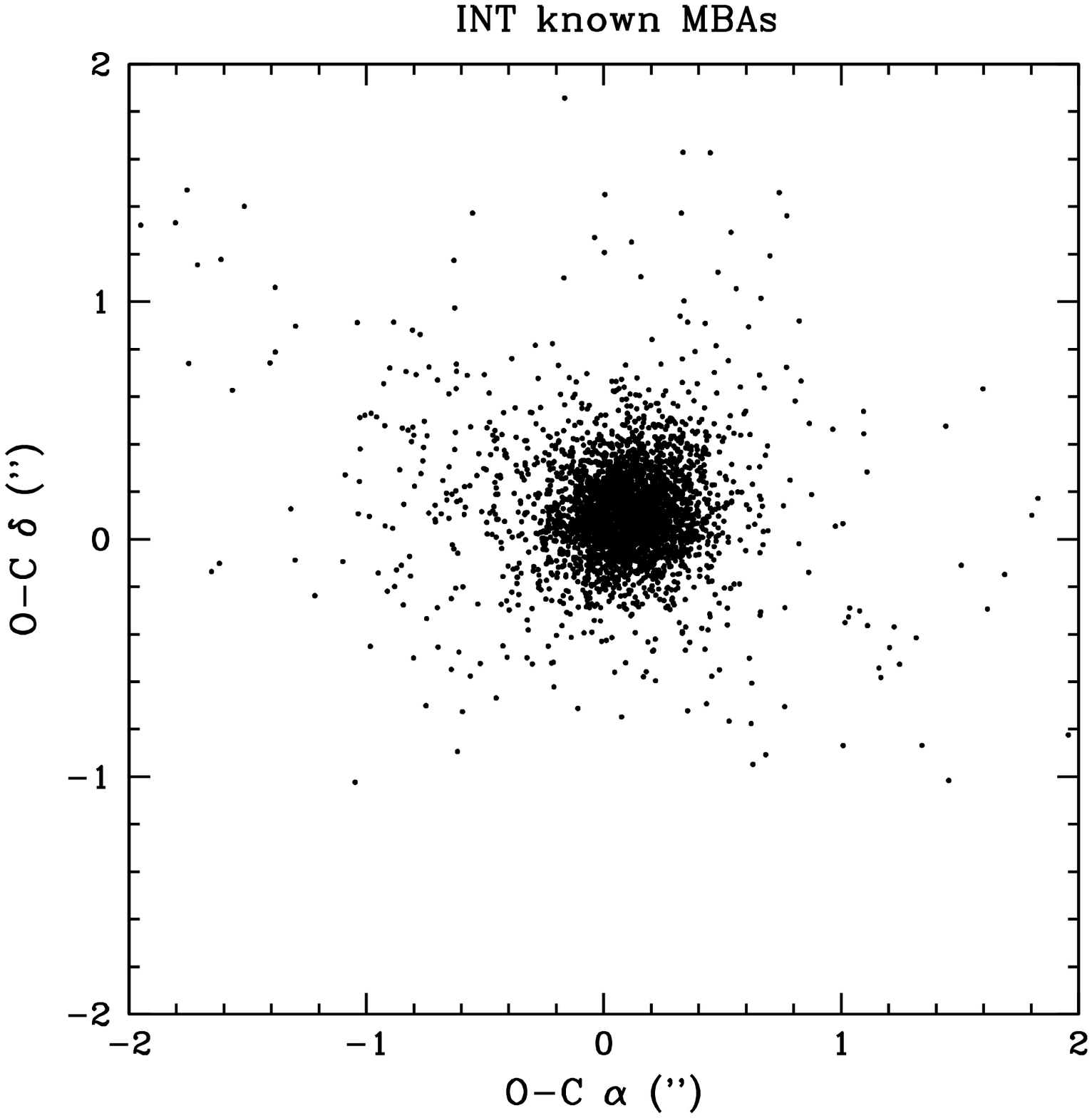}}
  }
  \caption{O--C (observed minus calculated) residuals for known MBAs observed with Blanco and INT. 
           INT-WFC images were corrected with THELI for field distortion, decreasing residuals across 
           the whole field to RMS standard deviation $0.41^{\prime\prime}$. }
  \label{fig3}
\end{figure*}

In Figure~\ref{fig3} we plot the O--C astrometric residuals corresponding to our reported 
positions of the known MBAs encountered in all Blanco and INT fields. The first panel plots 
the residuals for the Blanco run, clearly affected by the field distortion of the MOSAIC II 
camera, with a root mean square of the O--C residuals $0.90^{\prime\prime}$ and some values 
larger than $2^{\prime\prime}$ for the objects measured far from the center of the camera. 
In the second panel we plot the residuals for the INT-WFC fields observed during Dec 2011 
and Feb 2012 runs (3,465 positions). All the INT-WFC images were reduced with THELI software 
\citep{erb05} which corrected the field distortion across the whole WFC field. The INT 
(right) O--C plot in Figure~\ref{fig3} shows the accuracy of the WFC astrometry improved 
after field correction (RMS of O--C residuals $0.41^{\prime\prime}$). By comparison, the 
similar INT-WFC plot in our previous paper (Figure 4b from \cite{vad11}) includes WFC data 
processed without THELI and shows an RMS more than double around the origin ($0.97^{\prime\prime}$). 

Some inhomogeneity could be observed comparing the INT NEA O--C plot (Figure~\ref{fig2}) 
with the INT MBA plot (Figure~\ref{fig3}), namely the NEA plot seems more elongated towards 
the $\alpha$ direction while the MBA plot is more symmetric in both $\alpha$ and $\delta$. 
This is explained by the fact that residuals for NEAs are larger than for MBAs due 
to the measurement errors caused by NEAs moving faster in $\alpha$ than in $\delta$. 

\subsection{Unknown objects}
\label{unknown}

Working in a team, we could detect visually and measure all moving objects appearing in all fields 
observed with the INT and Blanco telescopes. All CCD images were bias corrected and flat fielded 
using IRAF or other image processing software, and only the WFC images were corrected for field distortion 
with THELI software \citep{erb05}. 
We used Astrometrica \citep{raa12} to identify all moving sources in the WFC and MOSAIC II paired CCDs 
based on the MPCORB asteroid database retrieved soon after the observing date. We measured and reported 
all the unknown objects which were later characterized by
determining preliminary orbits using the FIND\_ORB 
software \citep{gra12} in batch mode. Their main orbital parameters ($a$, $e$, $i$ and MOID) are given 
in the Appendix Tables A3-A7. Based on the observed fields and runs, we
distinguish the following lists which total 1,090 previously unknown asteroids: 

\begin{itemize}
\item 104 unknown objects discovered mostly during multiple nights in the Feb 2012 WFC opposition fields (Table A3); 
\item 626 unknown objects observed with the INT during one night in Feb 2012 in the opposition mini-survey fields (Table A4); 
\item 89 unknown objects observed with the INT in Feb 2012 in program NEA fields (Table A5); 
\item 75 unknown objects observed with the INT during 2011 runs in program NEA fields (Table A6); 
\item 196 unknown objects observed with Blanco telescope during one night in 2011 in program NEA fields (Table A7). 
\end{itemize}

\subsubsection{Discovered MBAs}

104 unknown objects could be recovered during multiple nights in the Feb 2012
WFC opposition mini-survey, thus they have been given provisional
designations and should soon become credited MBA discoveries of our EURONEAR
program using the INT WFC.  For these 104 discovered objects, Appendix Table
A3 lists firstly their orbital elements $a$, $e$, $i$ and MOID (calculated
with FIND\_ORB in batch mode) based on observational data from our run only
(linked by the MPC), and secondly their published MPC orbits based on all
available observations in the MPC database (accessed 8 Aug 2012). The
calculated FIND\_ORB orbits are similar to those published by the MPC for the
majority of our discovered MBAs, with standard deviations 0.26 AU in $a$,
0.14 in $e$, $3.78\deg$ in $i$ and 0.30 AU in MOID. In fact the MPC orbits
are calculated mostly using our data alone, but in some cases eventually
adding other observations from elsewhere before or following our runs.

\subsubsection{Unknown MBAs}

The great majority of the 1,090 unknown objects detected in the INT and Blanco runs could be 
characterised as MBAs, their proper motion and preliminary orbits matching the known main 
belt population well. We include the remaining unknown objects in Appendix Tables A4 to A7, listing 
orbital elements $a$, $e$ and $i$ obtained by FIND\_ORB automated fitting of 
our observational data. 
Due to the very short arcs derived from one night observations, the FIND\_ORB
preliminary orbits should be regarded with caution. Nevertheless, FIND\_ORB
was found to fit most orbits quite accurately, following direct comparison of
one night fitted orbits with published orbits based on all available
observations, and also after comparison with the whole known asteroid
population via the classic $a-e$ and $a-i$ plots (Figure 7 of \cite{vad11})
which can be virtually reproduced using the 1,090 preliminary orbits found
here.

\subsubsection{NEO candidates}
\label{cand}

\begin{figure*}
  \centerline{
    \mbox{\includegraphics[width=15cm]{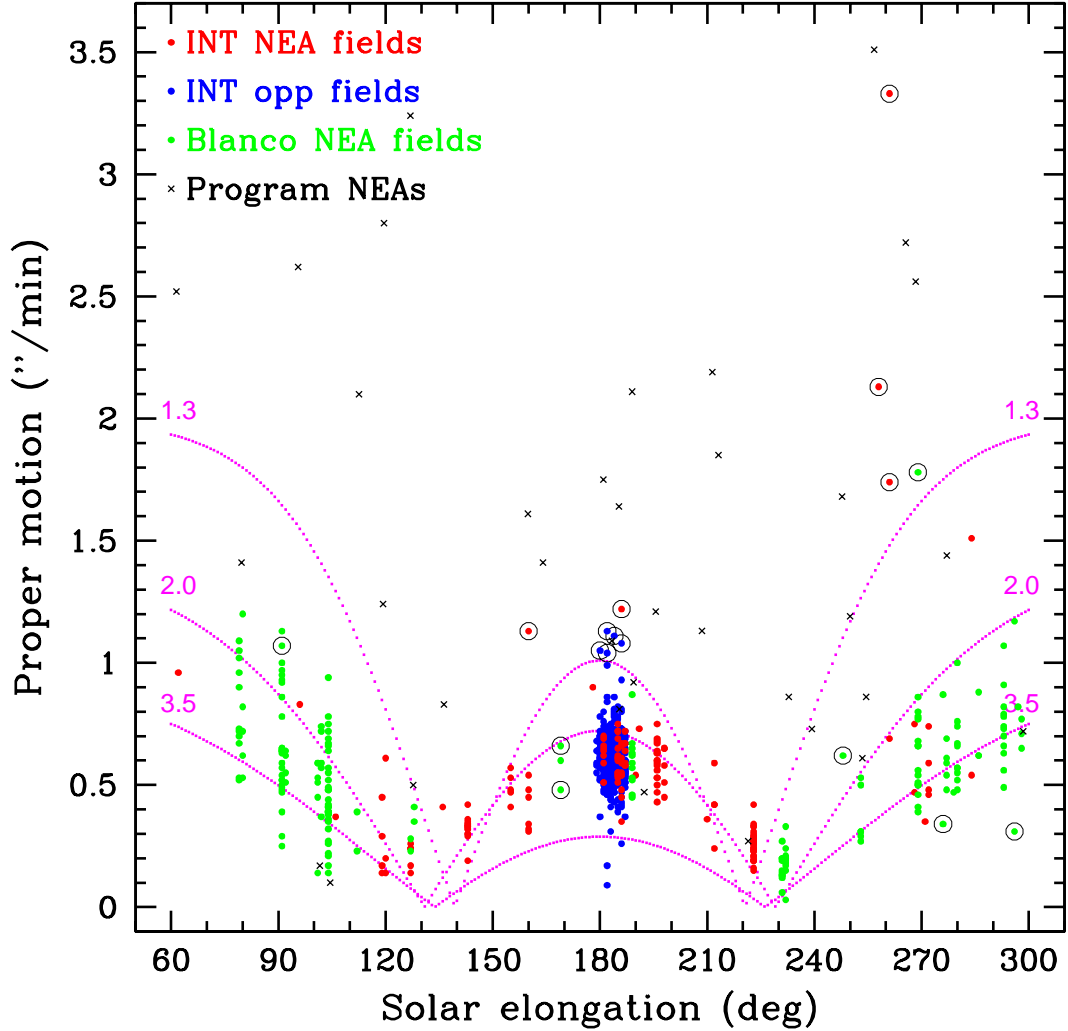}}
  }
  \caption{Using the $\epsilon-\mu$ orbital model \citep{vad11} to check NEO candidates based 
           on direct observational quantities Solar elongation ($\epsilon$) and proper motion ($\mu$). 
           We plot with solid symbols all unknown objects observed with Blanco (green) and the 
           INT (red for objects observed in program NEA fields and blue for objects observed in 
           opposition fields). 
           The three overlaid dotted magenta curves correspond to asteroids orbiting between 
           $a=2.0$ and $a=3.5$ AU (main belt) and $a=1.3$ AU (NEO limit). 
           We mark with circles the 18 best NEO candidates labeled in bold in the Appendix Table A8. 
           The fast NEO candidate ESU031 (identified with the known NEA 2012~DC28) is located above 
           the plot, matching the model. 
           With black crosses we plot the 47 program NEAs (which include 11 fast objects located 
           above the plot), most of them (38 objects or $81\%$) matching the $\epsilon-\mu$ model. 
} 
  \label{fig4}
\end{figure*}

Following our previous work which analyzed data taken with 1-2m telescopes \citep{vad11}, we 
use three independent search methods to check the unknown objects for Near Earth Objects 
(NEOs) observed with 2-4m facilities.
The first method employs our model presented by \cite{vad11} which plots the
two directly observed quantities $\mu$ (apparent proper motion) and
$\epsilon$ (Solar elongation, $\epsilon$ above 180 $\deg$ corresponding to
sky directions east of opposition).  Using this model (Figure~\ref{fig4}),
one can distinguish NEO candidates (located above the $a=1.3$ AU dotted
magenta curve corresponding to the NEO limit) from MBA objects (located below
the $a=2.0$ AU curve).  The last column of Appendix Table A8 lists the result
of this fit as ``Best'', ``Good'', ``Close'' or ``Bad'' with respect to this
model, considering only objects flagged ``Good'' or ``Best'' as NEO
candidates.

Our second NEO search method uses the ``NEO Rating'' server developed by the MPC \citep{mpc12b} which 
calculates a score for NEO candidates based on their expected proper motion
compared to the MBA distribution. 
In the second last column of Table A8 we include the MPC rating on a scale from $0\%$ (worst) 
to $100\%$ (best fit), considering objects having scores higher than $10\%$ as NEO candidates. 

Our third NEO search method uses the MOIDs of the preliminary orbits
calculated with FIND\_ORB. It should be regarded with caution because of the
small observed arcs, but can indicate possible NEOs.  We take MOID $<0.3$ AU
as the criterion for NEO candidates (Table A8).

From 1,090 unknown objects observed with INT and Blanco (Tables A3-A7), we
find 60 NEO candidates defined as satisfying at least {\it one} of the three
search methods (Table A8). To allow a safer selection, we highlight 18
{\it best NEO candidates} defined as unknown objects matching at least
{\it two} of the three selection methods (bold in Table A8 and marked with
circles in Figure~\ref{fig4}).

Five of these 18 best NEO candidates were observed
with the INT in 2012 and 2011 in program NEA fields:
EBA012, EPA015, VFLLP01, VKF008 and VKF034. 
The first three fulfilled all three NEO search criteria.
EBA012 was quite a fast moving object ($\mu=2.13^{\prime\prime}$/min)
observed in the field of NEA 2012~BB14 (recovered and published).
EPA015 was a fairly bright object ($R=19.6$) observed in the field of the
NEA 2011~FR29 (recovered and published). It too was among the fastest of the
best NEO candidates ($\mu=3.33^{\prime\prime}$/min), which classifies it as a
NEA.
VFLLP01 was a relatively faint object ($R=21.7$),
encountered at 12 Sep 2011 as the only object in the field of PHA 
2011~BN24 (not found at $V=21.6$), although the EURONEAR O--C calculator does not show any correlation 
between the two objects. Its FIND\_ORB fit is very unstable based on the least squares, Herget or 
downhill simplex methods, thus its calculated orbit should be regarded
with caution. 

Six best NEO candidates were observed with the INT during 2012 in the
opposition mini-survey: EBA023, EPA143, EPA190, EPO031, EPO065 and ESU031
(Table A8). ESU031 represents our fastest NEO candidate at
$\mu=10.32^{\prime\prime}$/min, producing long trails in three 180 sec
exposures taken in the opposition OP5 mini-survey field observed during the
first night 25 Feb 2012. We promptly reduced and submitted 3 positions to
MPC, but the robot matched it with 2012~DC28, an NEA discovered only two
days before by the Catalina survey.
EPO031 fulfils all three NEO selection criteria and has relatively high
brightness ($R=19.6$): the MPC linked it with an object seen by Catalina,
designating it 2012~DL54, but its orbit is still indeterminate.

The remaining seven best NEO candidates were observed with the Blanco
telescope: PCTV024, PCTV026, PCTVb50, PCV023, PCVP005, PCVP007 and PCVS024
(Table A8). The first two were encountered in the field of PHA 2008~YS27 (not
found). PCTV024 is the brightest ($R=18.3$) of our 18 best NEO candidates,
with MPC score $100\%$ and estimated MOID close to the NEA limit (0.28 AU).
PCTV026 is a very similar object measured only in 3 images. However, both
objects give a $\epsilon-\mu$ ``Bad'' flag, while their measured magnitudes
are highly uncertain (despite their brightness), being affected by the very
high Milky Way star density seen with the 4m telescope. PCV023 is a
relatively fast object $\mu=1.78^{\prime\prime}$/min, and as the only one of
the seven Blanco objects meeting all three NEO search citeria, is among the
most promising of all our best NEO candidates.

%% ==================================================================================================

\section{Discussion}
\label{disc}

\subsection{Comparison between NEO search methods}

At present (Oct 2012) 16 of the 18 best NEO candidates (Section \ref{cand})
remain as one-night objects, with one (2012~DC28) confirmed as a certain NEA.
We test the suitability of the three NEO search criteria by applying them to
objects known to be NEOs, namely the 47 program NEAs observed with INT and
Blanco. We test the search methods using our observed small arcs only
(knowing the objects to be NEOs from the MPC published orbits based on all
available observations).

To test the $\epsilon-\mu$ model, in Figure~\ref{fig4} we plot with crosses 
all 47 program NEAs. 38 objects (81\%) are confined above the NEO limit, while 
9 outliers have very elliptical orbits, namely 2008~QT3 ($e=0.52$), 2011~AG5 
($e=0.39$) and 2010~XC25 ($e=0.53$) at bottom-left of the plot, 2011~XZ1 
($e=0.46$) and 2009~OG ($e=0.86$) close to opposition, and 2008~XB1 ($e=0.37$), 
(175706) ($e=0.35$), (175189) ($e=0.39$) and 2007~JF22 ($e=0.59$) at 
bottom-right of the plot. 

Using the same 47 program NEA sets of observations, the NEO rating tool 
has 36 of them ($77\%$) satisfying the recommended MPC $Int>50\%$ limit 
\citep{mpc12b}. Of the remaining 11 objects, 8 also fail the $\epsilon-\mu$ 
model, namely: 2008~QT3 ($Int$ score $15\%$), 2010~XC25 ($25\%$), 2011~XZ1 
($3\%$), 2009~OG ($17\%$), 2008~XB1 ($20\%$), (175706) (score $23\%$), 
(175189) ($27\%$) and 2007~JF22 (score $4\%$). Three other outliers are 
2009~EE81 (10\%), 2006~CT10 ($16\%$) and 2012~AC13 (rate $47\%$). Only two 
objects have a NEO score below $10\%$, supporting this limit as a safe 
threshold for our selection method of NEO candidates.

Running FIND\_ORB for the same 47 program NEAs, using only our short observed
arcs, MOIDs of 11 objects result larger than the NEO limit of 0.3 AU,
i.e.\ 36 (76\%) would be NEOs according to this test. Nevertheless, comparing
these estimated MOIDs to the more accurately known MOIDs from MPC published
orbits based on all available observations, the standard deviations are quite
large, namely 0.90 AU in $a$, 0.34 in $e$, $12.9\deg$ in $i$ and 0.28 AU in
the MOID, thus the values of the FIND\_ORB fitted orbits of NEAs based on
very small arcs should be regarded with great caution.

In comparison the same three search criteria applied to our unknown objects
resulted in 13 objects as NEO candidates according to the $\epsilon-\mu$
model (flags ``Best'' or ``Good''), 26 objects according to the MPC ``NEO
rating'' selection ($Int$ score $> 10\%$), and 41 objects according to
their FIND\_ORB fitted orbits (MOID$<0.3$ AU), with 18 best NEO candidates
satisfying at least two of these criteria (Section \ref{cand}). Based on
this data sample, the $\epsilon-\mu$ model closely followed by the NEO
Rating appear the safest methods to flag one-night NEO candidates.

\subsection{Comparison between 2m and 4m facilities}
\label{comp}

The Blanco 4m run consisted of only one dark clear night, partially affected
by thin clouds. The INT runs consisted of about four mostly dark clear nights
in all, three during the Feb 2012 run, and time totalling about one night
during four nights in 2011. Thus, the INT 2.5m data could be more
representative statistically than the Blanco 4m data.

\begin{figure*}
  \centerline{
    \mbox{\includegraphics[width=7cm]{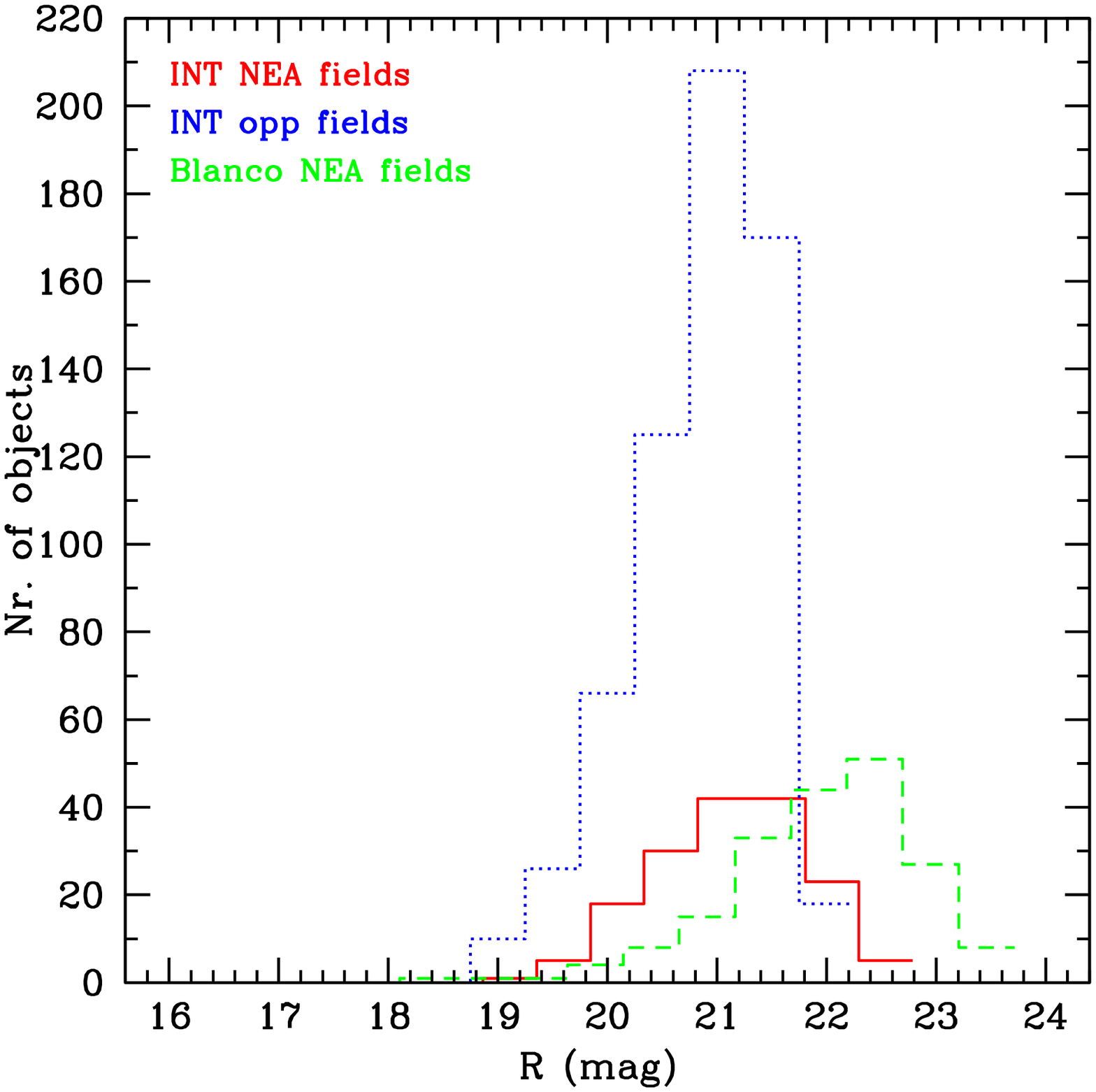}}
    \mbox{\includegraphics[width=7cm]{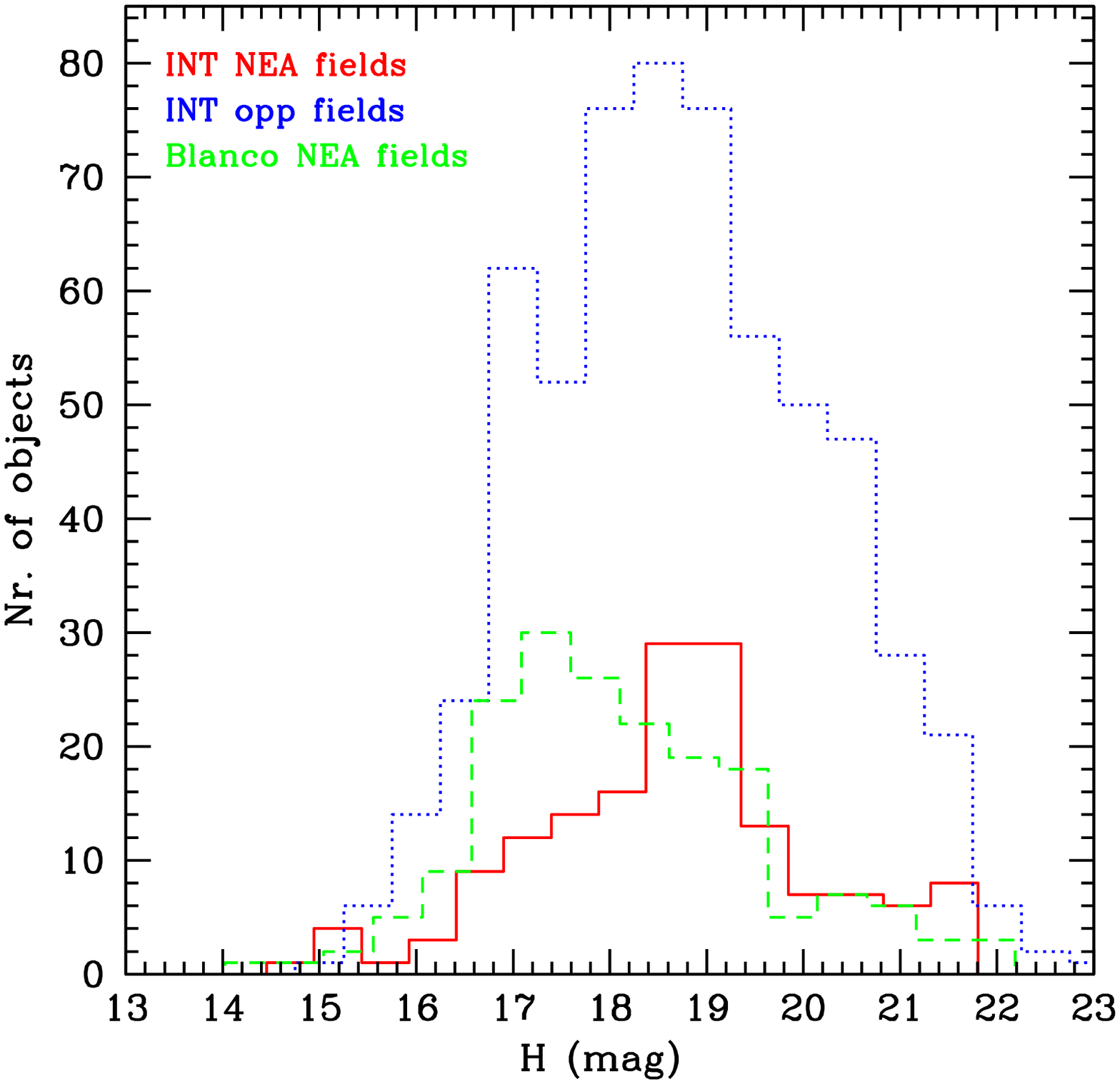}}
  }
  \caption{Number of unknown objects as a function of observed apparent $R$ magnitude 
          and calculated absolute magnitude $H$ for the INT dataset (red solid line 
          for NEA fields, blue dotted line for opposition fields) and Blanco (green dashed 
          line for NEA fields). } 
  \label{fig5}
\end{figure*}

In Figure~\ref{fig5} we plot histograms counting the unknown objects as a
function of the observed $R$ magnitude (left) and calculated absolute
magnitude $H$ (right) based on the three available datasets: the INT NEA
fields (2011 and 2012), INT opposition fields (2012), and Blanco NEA fields.
According to the left plot, the Blanco telescope sampled about 2 mag 
deeper than the INT, being most efficient in detection around $R\sim22.5$ and reaching a limit 
of $R\sim24$. The INT was most efficient around $R\sim21$ and reached a limit $R\sim22$. 
Based on the left tail of the total distribution, apparently there are no more unknown 
objects brighter than $R\sim19$ to discover nowadays. 
The INT limit is higher by about $0.8$ mag than our past results ($R\sim21.2$ in \cite{vad11}), 
probably due to better weather conditions in 2011-2012. Also, the INT limit appears to 
surpass by about 0.5 mag our past dark time ESO/WFI limit ($R\sim21.5$). 
The $H$ histogram suggests the Blanco telescope is finding a higher
proportion of larger (brighter $H$) objects than the INT, presumably because
many of these unknown objects are at larger distances (thus fainter $R$),
i.e.\ Blanco tends to see further out into the main belt.

Table~\ref{table2} summarizes data derived from Blanco and INT runs,
in observed NEA fields and opposition (opp.) fields. 
Based on the O--C residuals, the astrometric quality for program NEAs observed mostly in the 
center of the fields is identical for the two cameras, with a root mean square of the O--C residuals 
of $0.46^{\prime\prime}$. Nevertheless, the astrometry of the known MBAs recovered across the whole 
MOSAIC II field is more than two times worse (RMS $0.97^{\prime\prime}$) due to the uncorrected field 
of Blanco at its prime focus, compared with the INT WFC (RMS $0.41^{\prime\prime}$). 

%%\newpage
\begin{table}
\begin{center}
\caption{Summary and statistics based on the Blanco and INT runs. } 
\label{table2}
{\small
\begin{tabular}{lrrr}
\noalign{\smallskip}
\noalign{\smallskip}
\hline
\hline
\noalign{\smallskip}
\smallskip
Observations &  Blanco  &  INT  &  Total  \\
\hline
\noalign{\smallskip}
\noalign{\smallskip}
Recovered NEAs                                &    14   &    33  &    47  \\
Nr. of positions                              &    94   &   221  &   315  \\
\noalign{\smallskip}
\noalign{\smallskip}
O--C standard deviation ($^{\prime\prime}$)   &         &        &        \\
... program NEAs                              &  0.46   &  0.46  &   ---  \\
... known MBAs                                &  0.97   &  0.41  &   ---  \\
\noalign{\smallskip}
\noalign{\smallskip}
Known MBAs                                    &   288   &   580  &   868  \\
... in NEA fields                             &   288   &   132  &   420  \\
... in opp. fields                            &   ---   &   448  &   448  \\
Nr. of positions                              & 1,699   & 3,465  & 5,164  \\
\noalign{\smallskip}
\noalign{\smallskip}
Unknown objects                               &   196   &   790  &   986  \\
... in NEA fields                             &   196   &   164  &   360  \\
... in opp. fields                            &   ---   &   626  &   626  \\
Nr. of positions                              & 1,301   & 4,025  & 5,326  \\
\noalign{\smallskip}
\noalign{\smallskip}
Discovered objects                            &   ---   &   104  &   104  \\ 
Nr. of positions                              &   ---   & 1,153  & 1,153  \\
\noalign{\smallskip}
\noalign{\smallskip}
Nr. of observing nights                       &     1   &     4  &     5  \\
Nr. of program NEA fields                     &    28   &    47  &    75  \\
Nr. of opp. fields                            &   ---   &    40  &    40  \\
Total nr. of fields                           &    28   &    87  &   115  \\
Total nr. of CCD images                       &   224   &   348  &   572  \\
\noalign{\smallskip}
\noalign{\smallskip}
Sky coverage (sq.deg)                         &    10   &    24  &    34  \\
... NEA fields (sq.deg)                       &    10   &    13  &    23  \\
... opp. fields (sq.deg)                      &   ---   &    11  &    11  \\
\noalign{\smallskip}
\noalign{\smallskip}
Limiting magnitude ($R$)                      &  24.0   &  22.0  &   ---  \\
Apparent magnitude peak ($R$)                 &  22.5   &  21.0  &   ---  \\
\noalign{\smallskip}
\noalign{\smallskip}
Total nr. of objects                          &   498   & 1,507  &  2,005 \\
Total nr. of positions                        & 3,094   & 8,864  & 11,958 \\
\noalign{\smallskip}
\noalign{\smallskip}
Known MBA density (obj/sq.deg)                &         &        &        \\
... in NEA fields                             &    29   &    10  &   ---  \\
... in opp. fields                            &   ---   &    41  &   ---  \\
Unknown MBA density (obj/sq.deg)              &         &        &        \\
... in NEA fields                             &    19   &    13  &   ---  \\
... in opp. fields                            &   ---   &    66  &   ---  \\
Unknown NEO density (obj/sq.deg)              &   0.7   &   0.5  &   ---  \\
\noalign{\smallskip}
\noalign{\smallskip}
\hline
\hline
\end{tabular}
}
\end{center}
\end{table}

Counting the unknown and known asteroids (mostly MBAs) encountered in the
Blanco and INT fields, we could assess the ratio of the density of unknown to
known MBAs in any 4m and 2m class survey. The 87 observed WFC fields give
this ratio for the INT as 1.6 in opposition fields and 1.2 in program NEA
fields, thus an average of 1.4. This fraction is very consistent with the 1.3
result based on ESO/MPG data from our previous work and is higher than our
past 0.8 result based on previous INT data affected by bad weather
\citep{vad11}. Counting the unknown and known objects observed with Blanco
during one night in 28 fields, the ratio is 0.7 for Blanco, unexpectedly
smaller than that from the INT. This inconsistency could be explained due to
poor statistics affected by weather and field selection. During the first
part of the only Blanco night the clouds and cirrus strongly affected 9
fields from the total of 28 observed. Four other fields were in the Milky Way
and could not be scanned for objects (other than the program NEAs) due to the
extreme star density there. Also, seven other fields were observed at high
ecliptic latitude and could not serve for statistics. Eliminating all these
fields, we selected five fields with limiting detection magnitude $R\sim23$
(2008~XB1, 2008~DJ, 2008~EP6, 2010~XC25 and 2009~CS), more suitable for
statistics. Based on this selection, we derive the ratio 2.7 for Blanco in
relatively good weather conditions. In conclusion, Blanco could discover
about twice as many unknown MBAs with respect to known MBAs (ratio 2.7)
compared with the INT (ratio 1.4), assuming similar good weather and dark sky
conditions for both facilities.

\subsection{Statistics of the unknown MBA and NEO population}
\label{stats}

Next we derive some statistics based on the data obtained from 10 square degrees covered 
by Blanco and 24 square degrees covered by the INT (11 square degrees near opposition 
and 13 square degrees in random NEA fields). 

\subsubsection{Unknown MBA density}

An average of 29 known MBAs and 20 unknown MBAs per square degree could be observed with Blanco 
based on our small analyzed sample (mostly far from ecliptic) partially affected by poor weather. 
The best two fields observed in good conditions in the ecliptic (2010~XC25 and 2009~CS, both 
far from opposition) give similar densities to each other and an average of 45 known MBAs 
and 90 unknown MBAs per square degree, thus a total of {\it 135 MBAs per square degree visible 
with the Blanco 4m telescope (two unknown to every one unknown asteroid)} up to $R\sim23$. 
This density based on small non-opposition statistics is comparable with the work of \cite{yos03} 
who counted 208 MBAs per square degree up to $R\sim23.0$ based on the opposition SMBAS-I survey 
and 182 MBAs per square degree to the same limit from the SMBAS-II survey \citep{yos07}. 

Based on our larger INT dataset, 41 known and 66 unknown MBAs per square degree could be 
detected in average in the observed fields. Thus, a total of {\it 107 MBAs per square degree 
(1.6 unknown to every 1 known objects) could be detected at opposition by the INT} up to $R\sim22.0$. 
This density is comparable with our previous result of 63 MBAs per square degree (27 known, 
36 unknown) based on a similar survey using ESO/MPG \citep{vad11} up to a limit 0.5 mag shallower. 
The SMBAS surveys when counted to the same $R=22.0$ limit give 128 objects per square degree 
\citep{yos03} and 103 objects per square degree \citep{yos07}, very close to our present INT result. 

\subsubsection{Unknown NEA density}

Considering the best NEO candidates marked in bold in the Appendix Table A8, we estimate the 
unknown NEO density detectable in 2m and 4m surveys. Counting six NEO candidates in the INT NEA 
fields covering 13 square degrees, we derive an average of {\it 0.5 NEOs per square degree observable 
with the INT}. This is in very good agreement with five best NEO candidates observed in the INT 
opposition fields covering 11 square degrees and also within the range of our previous results 
(between 0.2 and 0.6 objects per square degree based on the ESO/MPG and from 0.1 to 0.8
for the INT, \cite{vad11}). 
In the Blanco program NEA fields covering 10 square degrees, we count seven
best NEO candidates, thus {\it 0.7 NEOs per square degree observable with the Blanco telescope}; 
this number should be regarded as a minimum, based on poor weather during the only available Blanco 
night. 

The MPC lists 609,956 known asteroids and 9,242 known NEAs in its database (checked 25 Oct 2012), 
a ratio of 1 in 66. Our NEO candidate list (Table A8) comprises 60 NEO candidates from which only 
18 are the best NEO candidates. Taking into account our total of 1,090 unknown objects observed 
with Blanco and INT, we derive a ratio of one NEO candidate for every 18 asteroids and 
{\it one best NEO candidate for every 60 unknown asteroids}. This ratio comes very close to the 
above known MPC ratio of 66, validating our NEO candidate selection method. 

\subsubsection{Total number of NEAs observable by a 2m survey}

Based on the 11 best NEO candidates discovered in 24 square degrees surveyed with the INT, 
we can evaluate the total number of unknown NEOs detectable by a 2m-class telescope. 
Checking the orbital distribution of our NEO candidates (11 INT objects in bold in the 
Appendix Table A.8) we observe that all these objects have inclination $i < 25^\circ$ 
(although the uncertainties are expected to be large due to small observed arcs). 
Considering the whole known NEA population (NEODyS 2012), we derive $86\%$ NEAs having 
$i < 25^\circ$, so we adopt the same percentage for our statistical analysis.
The area of the celestial sphere between ecliptic latitudes $-25^\circ$ and $+25^\circ$ 
represents $17,400$ square degrees, namely about $42\%$ from the total area of a sphere. 
Extrapolating our number of 11 objects observed in 24 square degrees within the ecliptic zone 
$-25^\circ < \beta < +25^\circ$, we derive about 8,000 unknown best NEO candidates having 
$i < 25^\circ$. Extrapolating in turn this number to all inclinations, we conclude that there 
are in total some 9,300 unknown NEO candidates observable with a 2m survey. 
Adding this number to the currently known almost 10,000 NEAs (discovered by 1-2m surveys), we 
predict about 19,300 NEAs observable by 2m surveys. This number is very close to \cite{mai12} 
who used NEOWISE data to predict a total of $20,500 \pm 4,200$ NEAs larger than 100m, which 
could be regarded as a limit for 2m surveys.

%% ==================================================================================================

\section{Conclusions and future work}
\label{future}

In this paper we report on the follow-up and recovery of 477 program NEAs, PHAs and VIs observed 
mostly between 2010 and 2012 from nine sites of the EURONEAR network which include six professional 
1-4m class telescopes located in Chile, La Palma, France and Germany plus three smaller educational 
and public outreach telescopes in Romania and Germany. The addition of these objects to our 
previous work leads to a total of 739 NEAs followed-up by the EURONEAR network since 2006. 

The reduced data presented in this paper generated 98 MPC publications. 
We use the most important data to study the orbit improvement of 111 NEAs, PHAs and VIs, 
among which 29 objects were recovered at a new opposition and the other 82 were followed-up 
soon after discovery. Although recovery of NEAs a few years after their last observation appears to 
be important, follow-up of newly discovered objects soon after discovery seems to be more valuable 
for the recovery and orbital improvement of fainter and highly uncertain objects which could remain 
invisible to existing surveys. 

We characterize each site based on the astrometric residual plot for all observed NEAs, finding 
WHT-ACAM, INT-WFC and Blanco-MOSAIC II (close to center only) the best instruments (RMS of the O--C
$0.44^{\prime\prime}-0.46^{\prime\prime}$), followed by Bonn and Galati ($0.56^{\prime\prime}-0.57^{\prime\prime}$), 
particularily good small educational facilities. Using our published data of known MBAs, we compare the astrometry 
across the large field of the INT-WFC and Blanco-MOSAIC II and also the astrometric improvement 
over the whole field of the INT-WFC after image correction of its quite distorted prime focus 
field, reaching RMS $0.41^{\prime\prime}$, more than two times better than $0.97^{\prime\prime}$, 
our past results without THELI correction. 

During two runs plus a few other discretionary hours
(equivalent to 5 clear nights in total) we used the 
large field INT-WFC and Blanco-MOSAIC II facilities to recover and follow-up 47 highly uncertain 
NEAs and also to carry out an opposition mini-survey with the INT (10 square degrees) focused on MBAs. 
In 115 fields observed with these two 2-4m large field facilities covering a
total of 34 sky square degrees we carefully measured and reported to the MPC,
in addition to the 47 program NEAs, all identified moving objects, comprising
868 known MBAs, 986 unknown objects - mostly MBAs, and 104 newly 
discovered MBAs. We use these data to derive new MBA and NEA observability statistics for 2-4m 
surveys, continuing our previous work based on data obtained using 1-2m large field facilities. 
The INT or any similar 2m-class telescope can observe NEAs as faint as $R\sim22$, being most 
efficient in discovery of MBAs around $R\sim21$, while Blanco or a similar 4m class facility 
can detect NEAs up to $R\sim24$, being most efficient to discover new objects around $R\sim22.5$. 
Based on our INT dataset, a total of 107 MBAs per square degree can be detected at opposition 
up to $R\sim22.0$ in a 2m survey (this density surpassing our 2m past results but remaining very 
close to other results based on surveys using larger telescopes). The two best fields observed 
with Blanco in the ecliptic in good weather conditions result in 135 MBAs per
square degree to $R\sim23$.

The ratio of unknown to known MBAs observable by a 2m or 4m class survey is
1.4 based on our INT data (very consistent with our past work) and 2.7 based
on a few fields observed in good weather with Blanco, which could discover
twice as many unknown MBAs with respect to known MBAs than the INT.

104 new MBAs were recovered in two or three nights during our opposition mini-survey using the INT, 
so we expect them to be credited in the future as EURONEAR MBA discoveries. 
We studied our 986 one-night unknown objects observed with Blanco and the INT using three 
independent NEO search criteria, finding 60 NEO candidates satisfying
at least one criterion and 18 best NEO candidates satisfying at least two. 
Using our total sky coverage (24 square degrees 
with INT and 10 square degrees with Blanco) and counting the best NEO candidates, we derived 
an average of 0.5 NEO candidates per square degree observable in a 2m survey (in very good 
agreement with our past results) and at least 0.7 NEO candidates per square degree based on our 
one Blanco night partially affected by clouds. The ratio of unknown objects
over best NEO candidates observed with INT and Blanco is 60, in very good
agreement with the ratio of 66 obtained when using the entire known published
asteroid and NEA populations. {\it Using the 11 best NEO candidates observed with the INT 
in 24 square degrees and a simple two step orbital model, we assess the total number of NEAs 
detectable by a 2m survey to 19,300 objects, in very close agreement to a recent work. }

We conclude by listing two future projects aiming to continue the study of the 
NEA distribution at the faint end, using data from existing large field 4-8m class telescopes. 
The first facility is the new DECam camera installed on the Blanco telescope. Second
is the SuprimeCam camera on Subaru (based on archival data related to another EURONEAR 
data-mining project) and also the new Subaru Hyper-SuprimeCam camera, which
could yield extensive statistics allowing existing NEO models to be checked
and the formation and evolution of the entire NEA population to be studied.
These facilities could be scanned using automated detection software, then
comparing results with human search based on our experience and other
independent studies.

%% ==================================================================================================

\section{Acknowledgements}

Special thanks are due to the science committees and institutions awarding time to our EURONEAR 
proposals: the Chilean National Time Committee (for the Blanco observing run 0646, 3-4 Jun 2011) 
and the Spanish Time Allocation Committee (for the INT observing run C6, 25-28 Feb 2012). 
Special thanks are due to the Isaac Newton Group (ING), the U.K. Science and Technology Facilities 
Council (STFC), the Spanish ``Ministerio de Ciencia e Innovaci\'on'' (MICINN) and the ``Instituto 
Astrofisico de Canarias'' (IAC) for supporting the project AYA2008-06202-C03-02 thanks to which 
six visiting students received funding for their INT observing run. M. Karami thanks the Iranian 
National Observatory project for the INT training opportunity and the INT observing nights. 
To correct the optical field distortions of the INT-WFC images we used THELI software; thanks are 
due to M. Schirmer for his assistance in using it. 
Acknowledgements are due to Bill Gray for providing, developing and allowing free usage of FIND\_ORB 
to the entire amateur-professional community. 
We thank the Minor Planet Center, especially T. Spahr and G. Williams who revised our MPC reports. 
Thanks are also due to the two referees who provided feedback important to improve our paper. 
This research has made intensive use of the Astrometrica software developed by Herbert 
Raab, very simple to install and use by students and amateur astronomers. 
We also used the image viewer SAOImage DS9, developed by Smithsonian Astrophysical 
Observatory and also IRAF, distributed by the National Optical Astronomy Observatories, 
operated by the Association of Universities for Research in Astronomy, Inc.\ under cooperative 
agreement with the National Science Foundation. 

\bibliographystyle{elsarticle}

\end{document}

\textheight = 24.5cm
\voffset=-0.8in
\hoffset=-0.5in

\appendix
\section{Data Tables - 739 observed NEAs and new 2-4m survey statistics within the EURONEAR network} 
\vspace{10 mm}
\subsection{The observing log for 477 program NEAs - page 2}
\vspace{5 mm}
\subsection{Comparison of orbits of 111 recovered NEAs - page 10}
\vspace{5 mm}
\subsection{104 unknown objects discovered mostly during multiple nights in the Feb 2012 WFC opposition fields - page 14}
\vspace{5 mm}
\subsection{626 unknown objects observed with the INT during one night in Feb 2012 in the opposition mini-survey fields - page 18}
\vspace{5 mm}
\subsection{89 unknown objects observed with the INT in Feb 2012 in program NEA fields - page 28}
\vspace{5 mm}
\subsection{75 unknown objects observed with the INT during 2011 runs in program NEA fields - page 30}
\vspace{5 mm}
\subsection{196 unknown objects observed with Blanco telescope during one 2011 night in program NEA fields - page 32}
\vspace{5 mm}
\subsection{60 NEO candidates and 18 best NEO candidates - page 35}

%%\documentclass[preprint,authoryear,12pt]{elsarticle}
%%\usepackage{longtable}

%%\begin{document}

% ______________________________

% APPENDIX TABLES
% ______________________________

% ESO/MPG DATA 
%__________

\newpage
%%\appendix

%%\addcontentsline{toc}{chapter}{Appendices}
%%\chapter{Appendix A: Program NEA objects observed with Blanco, INT, WHT, OHP 1.2m, Pic du Midi 1m, Tautenburg 2m, Bonn AIfA 0.5m, Galati 0.4m and Urseanu 0.3m telescopes}

%%\section{TITLU 1}
%%Program NEA objects observed with Blanco, INT, WHT, OHP 1.2m, Pic du Midi 1m, Tautenburg 2m, Bonn AIfA 0.5m, Galati 0.4m and Urseanu 0.3m telescopes} 

\scriptsize
\begin{center}
\begin{longtable}{llrrrrrrr} 
%Here is the caption, the stuff in [] is the table of contents entry,
%the stuff in {} is the title that will appear on the first page of the
%table.
\caption{ The observing log for 477 program NEAs. We list the name of the asteroid, 
its classification, date of observation, expected apparent magnitude $V$, exposure time (seconds), 
number of observed positions, apparent motion $\mu$ ($^{\prime\prime}$/min) and the length of the 
arc before and after our observations (d-days, m-months, y-years). Objects recovered at a new opposition 
are marked by $\star$ while new objects followed-up soon after discovery are marked by $\bullet$ } 
\\
%This is the header for the first page of the table...
\hline \hline \\
   \multicolumn{1}{c}{Telescope} &
   \multicolumn{1}{c}{Asteroid} &
   \multicolumn{1}{c}{Class} &
   \multicolumn{1}{c}{Date (UT)} &
   \multicolumn{1}{c}{$V$} &
   \multicolumn{1}{c}{Exp} &
   \multicolumn{1}{c}{Nr pos} &
   \multicolumn{1}{c}{$\mu$ ($^{\prime\prime}$/min)} &
   \multicolumn{1}{c}{Arc} \\
\\ \hline \hline \\
\endfirsthead

%This is the header for the remaining page(s) of the table...
\multicolumn{9}{c}{{\tablename} \thetable{} (continued) -- The observing log for 477 program NEAs. } \\
\\  \hline \hline \\
   \multicolumn{1}{c}{Telescope} &
   \multicolumn{1}{c}{Asteroid} &
   \multicolumn{1}{c}{Class} &
   \multicolumn{1}{c}{Date (UT)} &
   \multicolumn{1}{c}{$V$} &
   \multicolumn{1}{c}{Exp} &
   \multicolumn{1}{c}{Nr pos} &
   \multicolumn{1}{c}{$\mu$ ($^{\prime\prime}$/min)} &
   \multicolumn{1}{c}{Arc} \\
\\ \hline \hline \\
\endhead

Blanco            &   (175189) 2005 EC224    &   NEA   &  2011 Jun 04    &   21.0   &    20   &	  7   &   1.4  &   9y/9y   \\ 
  	          &   2008 XB1 $\star$       &   PHA   &  2011 Jun 04    &   23.3   &   120   &	  6   &   0.6  &   2m/3y   \\ 
  	          &   2010 KX7	             &   PHA   &  2011 Jun 04    &   21.0   &    20   &	  6   &   1.7  &   1y/1y   \\ 
  	          &   2009 BE81 $\star$      &   PHA   &  2011 Jun 04    &   23.6   &    90   &	  7   &   0.9  &   2m/2y   \\ 
  	          &   2009 OG $\star$        &   PHA   &  2011 Jun 04    &   23.4   &   120   &	  7   &   0.5  &   5m/2y   \\ 
  	          &   2003 VE1 $\star$       &   PHA   &  2011 Jun 04    &   21.2   &    20   &	  6   &   1.1  &   1m/8y   \\ 
  	          &   2006 CT10	$\star$      &   PHA   &  2011 Jun 04    &   23.3   &   120   &	  7   &   0.9  &   3m/5y   \\ 
  	          &   2009 BD	             &   NEA   &  2011 Jun 04    &   16.5   &     2   &	  6   &  36.0  &   2y/2y   \\
\end{longtable}
\end{center}

% _________

%%\end{document}

%%\documentclass[preprint,authoryear,12pt]{elsarticle}
%%\usepackage{longtable}

%%\begin{document}

% ______________________________

% APPENDIX TABLES
% ______________________________

% ESO/MPG DATA 
%__________

\newpage
%%\appendix

%%\addcontentsline{toc}{chapter}{Appendices}
%%\chapter{Appendix A: Program NEA objects observed with Blanco, INT, WHT, OHP 1.2m, Pic du Midi 1m, Tautenburg 2m, Bonn AIfA 0.5m, Galati 0.4m and Urseanu 0.3m telescopes}

%%\section{TITLU 1}
%%Program NEA objects observed with Blanco, INT, WHT, OHP 1.2m, Pic du Midi 1m, Tautenburg 2m, Bonn AIfA 0.5m, Galati 0.4m and Urseanu 0.3m telescopes} 

\scriptsize
\begin{center}
\begin{longtable}{lrrrrrrrrr} 
%Here is the caption, the stuff in [] is the table of contents entry,
%the stuff in {} is the title that will appear on the first page of the
%table.
\caption{ Comparison of orbits of 111 program NEAs. Objects recovered at a new opposition 
         are marked by $\star$ while new objects followed-up soon after discovery are marked by $\bullet$. 
         For every object, the first line gives orbital elements fitted without our data, while the second line gives elements adding our data. 
         Orbital elements fitted with FIND\_ORB (Gray, 2012) at epoch $JD=2456200.5$ listing the asteroid name, semimajor axis $a$, 
         eccentricity $e$, inclination $i$, longitude of the ascending node $\Omega$, argument of pericenter $\omega$, and mean anomaly $M$, 
         followed by the minimal orbital intersection distance MOID, number of fitted positions and the root mean square residual of the 
         fit $\sigma$. } 
\\
%This is the header for the first page of the table...
\hline \hline \\
   \multicolumn{1}{c}{Asteroid} &
   \multicolumn{1}{c}{$a$ (AU)} &
   \multicolumn{1}{c}{$e$} &
   \multicolumn{1}{c}{$i$ ($\degree$)} &
   \multicolumn{1}{c}{$\Omega$ ($\degree$)} &
   \multicolumn{1}{c}{$\omega$ ($\degree$)} &
   \multicolumn{1}{c}{$M$ ($\degree$)} &
   \multicolumn{1}{c}{MOID (AU)} &
   \multicolumn{1}{c}{Nr pos} &
   \multicolumn{1}{c}{ $\sigma$ ($^{\prime\prime}$)} \\
\\ \hline \hline \\
\endfirsthead

%This is the header for the remaining page(s) of the table...
\multicolumn{10}{c}{{\tablename} \thetable{} (continued) -- Comparison of orbits of 111 program NEAs. } \\
\\  \hline \hline \\
   \multicolumn{1}{c}{Asteroid} &
   \multicolumn{1}{c}{$a$ (AU)} &
   \multicolumn{1}{c}{$e$} &
   \multicolumn{1}{c}{$i$ ($\degree$)} &
   \multicolumn{1}{c}{$\Omega$ ($\degree$)} &
   \multicolumn{1}{c}{$\omega$ ($\degree$)} &
   \multicolumn{1}{c}{$M$ ($\degree$)} &
   \multicolumn{1}{c}{MOID (AU)} &
   \multicolumn{1}{c}{Nr pos} &
   \multicolumn{1}{c}{ $\sigma$ ($^{\prime\prime}$)} \\
\\ \hline \hline \\
\endhead

\multicolumn{10}{c}{Blanco observations: } \\
\noalign{\smallskip}\noalign{\smallskip}

2008 XB1 $\star$    &  1.3325316 &  0.3727211 &  12.68489 &  270.40512 &   84.00333 &  210.02902 &  0.0460 &  90 &  0.54 \\
                    &  1.3325334 &  0.3727214 &  12.68489 &  270.40513 &   84.00348 &  210.02716 &  0.0460 &  96 &  0.54 \\

2009 BE81 $\star$   &  1.5224083 &  0.3494415 &  18.79196 &  118.05393 &  333.38243 &  356.52156 &  0.0326 &  32 &  0.53 \\
                    &  1.5226071 &  0.3495222 &  18.79412 &  118.05268 &  333.38496 &  356.38013 &  0.0326 &  39 &  0.50 \\

2009 OG $\star$     &  2.7049543 &  0.8590749 &  48.36843 &  309.43967 &  114.15714 &  242.33456 &  0.0411 & 211 &  0.52 \\
                    &  2.7049503 &  0.8590747 &  48.36839 &  309.43968 &  114.15714 &  242.33509 &  0.0411 & 218 &  0.52 \\

2003 VE1 $\star$    &  1.9424421 &  0.4950035 &  16.31434 &   29.65747 &  323.10760 &  123.48639 &  0.0442 &  62 &  0.41 \\
\end{longtable}
\end{center}

% _________

%%\end{document}

%%\documentclass[preprint,authoryear,12pt]{elsarticle}
%%\usepackage{longtable}

%%\begin{document}

% ______________________________

% APPENDIX TABLES
% ______________________________

\newpage
%%\appendix
%%\section{104 unknown asteroids discovered at the INT in Feb 2012 in the opposition mini-survey. } 

\scriptsize
\begin{center}
\begin{longtable}{llrrrrrrrrrr}
%Here is the caption, the stuff in [] is the table of contents entry,
%the stuff in {} is the title that will appear on the first page of the
%table.
\caption{ 104 unknown MBAs discovered at the INT in Feb 2012 in the opposition mini-survey fields. 
2-night and 3-night observations of the same objects are linked and
designated by the MPC; their corresponding acronyms previously assigned
by us are listed in the first column. 
In the first line we give the orbital elements fitted with FIND\_ORB based on
linked observational 
data from our run only. In the second line we include published MPC orbital data based on all available 
observations, according to the MPC database accessed at 8 Aug 2012.} 
\\
%This is the header for the first page of the table...
\hline \hline \\
   \multicolumn{1}{c}{Acronym} &
   \multicolumn{1}{c}{Designation} &
   \multicolumn{1}{c}{$a$} &
   \multicolumn{1}{c}{$e$} &
   \multicolumn{1}{c}{$i$} &
   \multicolumn{1}{c}{MOID} &
   \multicolumn{1}{c}{$H$} &
   \multicolumn{1}{c}{Nr pos} &
   \multicolumn{1}{c}{Arc} &
   \multicolumn{1}{c}{$\sigma$} &
   \multicolumn{1}{c}{$R$} &
   \multicolumn{1}{c}{$\mu$} \\
   %% units bellow
   \multicolumn{1}{c}{} &
   \multicolumn{1}{c}{(from MPC)} &
   \multicolumn{1}{c}{(AU)} &
   \multicolumn{1}{c}{} &
   \multicolumn{1}{c}{($\degree$)} &
   \multicolumn{1}{c}{(AU)} &
   \multicolumn{1}{c}{} &
   \multicolumn{1}{c}{} &
   \multicolumn{1}{c}{} &
   \multicolumn{1}{c}{($^{\prime\prime}$)} &
   \multicolumn{1}{c}{} &
   \multicolumn{1}{c}{($^{\prime\prime}$/min)} \\
\\ \hline \hline \\
\endfirsthead

%This is the header for the remaining page(s) of the table...
\multicolumn{12}{c}{{\tablename} \thetable{} (continued) -- 104 unknown asteroids discovered at the INT in Feb 2012 in the opposition mini-survey. } \\
\\  \hline \hline \\
   \multicolumn{1}{c}{Acronym} &
   \multicolumn{1}{c}{Designation} &
   \multicolumn{1}{c}{$a$} &
   \multicolumn{1}{c}{$e$} &
   \multicolumn{1}{c}{$i$} &
   \multicolumn{1}{c}{MOID} &
   \multicolumn{1}{c}{$H$} &
   \multicolumn{1}{c}{Nr pos} &
   \multicolumn{1}{c}{Arc} &
   \multicolumn{1}{c}{$\sigma$} &
   \multicolumn{1}{c}{$R$} &
   \multicolumn{1}{c}{$\mu$} \\
   %% units bellow
   \multicolumn{1}{c}{} &
   \multicolumn{1}{c}{} &
   \multicolumn{1}{c}{(AU)} &
   \multicolumn{1}{c}{} &
   \multicolumn{1}{c}{($\degree$)} &
   \multicolumn{1}{c}{(AU)} &
   \multicolumn{1}{c}{} &
   \multicolumn{1}{c}{} &
   \multicolumn{1}{c}{} &
   \multicolumn{1}{c}{($^{\prime\prime}$)} &
   \multicolumn{1}{c}{} &
   \multicolumn{1}{c}{($^{\prime\prime}$/min)} \\
\\ \hline \hline \\
\endhead

\multicolumn{12}{c}{INT Feb 2012 3-night objects discovered in the opposition mini-survey: } \\
\noalign{\smallskip}\noalign{\smallskip}

EBA017b EBA091 EBA149 	&	2012 DS41	&	2.41	&	0.17	&	1.88	&	1.01	&	18.8	&	14	&	3d	&	0.17	&	21.4	&	0.64	\\
                      	&	           	&	2.51	&	0.31	&	2.37	&	0.74	&	18.3	&	14	&	3d	&	      	&	      	&	     	\\
EBA028 EBA100 EVA035  	&	2012 DM41	&	3.14	&	0.15	&	0.32	&	1.82	&	17.4	&	15	&	3d	&	0.17	&	21.3	&	0.51	\\
                      	&	           	&	3.21	&	0.11	&	0.27	&	      	&	17.9	&	15	&	3d	&	      	&	      	&	     	\\
EBA037 EBA101 EBA134  	&	2012 DT46	&	2.64	&	0.14	&	3.25	&	1.27	&	17.3	&	15	&	3d	&	0.12	&	20.6	&	0.59	\\
                      	&	           	&	2.62	&	0.12	&	3.27	&	      	&	17.2	&	19	&	3d	&	      	&	      	&	     	\\
EBA042 EBA105 EBA138  	&	2012 DR41	&	2.34	&	0.23	&	3.20	&	0.79	&	18.5	&	15	&	3d	&	0.11	&	21.1	&	0.68	\\
\end{longtable}
\end{center}

%%\end{document}

%%\documentclass[preprint,authoryear,12pt]{elsarticle}
%%\usepackage{longtable}

%\begin{document}
%%\setlength{\topmargin}{-0.8in}

% ______________________________

% APPENDIX TABLES
% ______________________________

% ESO/MPG DATA 
%__________

\newpage
%%\appendix
%%\section{Unknown objects observed during the Feb 2012 opposition mini-survey with the INT telescope. } 

\scriptsize
\begin{center}

\begin{longtable}{llrrrrrrrrrrp{2cm}}

%Here is the caption, the stuff in [] is the table of contents entry,
%the stuff in {} is the title that will appear on the first page of the
%table.
\caption{ 626 unknown objects observed with the INT during one night in Feb 2012 in the opposition mini-survey fields. 
We include approximate orbital data fitted with FIND\_ORB software based on our 1-night datasets.} 
\\
%This is the header for the first page of the table...
\hline \hline \\
%%\line(0,0){10 cm}
   \multicolumn{1}{c}{Acronym} & 
   \multicolumn{1}{c}{$a$ (AU)} &
   \multicolumn{1}{c}{$e$} &
   \multicolumn{1}{c}{$i$ ($\degree$)} &
   \multicolumn{1}{c}{MOID (AU)} &
   \multicolumn{1}{c}{$H$} &
   \multicolumn{1}{c}{Nr pos} &
   \multicolumn{1}{c}{Arc} &
   \multicolumn{1}{c}{$\sigma$ ($^{\prime\prime}$)} &
   \multicolumn{1}{c}{$R$} &
   \multicolumn{1}{c}{$\beta$ ($\degree$)} &
   \multicolumn{1}{c}{$\epsilon$ ($\degree$)} &
   \multicolumn{1}{p{2cm}}{$\mu$ ($^{\prime\prime}$/min)} \\
\\ \hline \hline \\
\endfirsthead

%This is the header for the remaining page(s) of the table...
\multicolumn{13}{c}{{\tablename} \thetable{} (continued) -- 626 unknown objects observed with the INT during one night in Feb 2012 in the opposition mini-survey fields. } \\
\\  \hline \hline \\
   \multicolumn{1}{c}{Acronym} & 
   \multicolumn{1}{c}{$a$ (AU)} &
   \multicolumn{1}{c}{$e$} &
   \multicolumn{1}{c}{$i$ ($\degree$)} &
   \multicolumn{1}{c}{MOID (AU)} &
   \multicolumn{1}{c}{$H$} &
   \multicolumn{1}{c}{Nr pos} &
   \multicolumn{1}{c}{Arc} &
   \multicolumn{1}{c}{$\sigma$ ($^{\prime\prime}$)} &
   \multicolumn{1}{c}{$R$} &
   \multicolumn{1}{c}{$\beta$ ($\degree$)} &
   \multicolumn{1}{c}{$\epsilon$ ($\degree$)} &
   \multicolumn{1}{p{2cm}}{$\mu$ ($^{\prime\prime}$/min)} \\
\\ \hline \hline \\
\endhead

EBA016	&	2.57	&	0.01	&	2.3	&	1.55	&	19.2	&	5	&	1n	&	0.16	&	21.8	&	-1	&	184	&	0.62	\\
EBA022	&	2.69	&	0.06	&	3.1	&	1.52	&	17.1	&	5	&	1n	&	0.09	&	20.4	&	-1	&	184	&	0.59	\\
EBA023	&	1.04	&	0.04	&	0.4	&	0.01	&	32.0	&	5	&	1n	&	0.27	&	20.7	&	-1	&	184	&	1.11	\\
EBA024	&	1.03	&	0.11	&	0.2	&	0.01	&	36.5	&	4	&	1n	&	0.38	&	20.4	&	-1	&	184	&	0.80	\\
EBA027	&	2.33	&	0.29	&	3.7	&	0.67	&	21.5	&	4	&	1n	&	0.09	&	21.5	&	0	&	184	&	0.57	\\
EBA029	&	2.50	&	0.24	&	2.6	&	0.90	&	19.5	&	5	&	1n	&	0.11	&	20.4	&	0	&	184	&	0.58	\\
EBA030	&	2.93	&	0.03	&	1.1	&	1.86	&	15.7	&	5	&	1n	&	0.05	&	19.0	&	0	&	184	&	0.56	\\
EBA031	&	2.30	&	0.30	&	0.1	&	0.63	&	21.4	&	5	&	1n	&	0.06	&	21.2	&	0	&	184	&	0.50	\\
\end{longtable}
\end{center}

%%\end{document}

%%\documentclass[preprint,authoryear,12pt]{elsarticle}
%%\usepackage{longtable}

%%\begin{document}
%%\setlength{\topmargin}{-0.8in}

% ______________________________

% APPENDIX TABLES
% ______________________________

% ESO/MPG DATA 
%__________

\newpage
%%\appendix
%%\section{Unknown objects observed with the INT during Feb 2012 in program NEA fields. } 

\scriptsize
\begin{center}

\begin{longtable}{llrrrrrrrrrrp{2cm}}
%%\begin{longtable}{*{13}{>{\centering\arraybackslash}c}}
%%\begin{longtable}{ccccccccccccc}

%Here is the caption, the stuff in [] is the table of contents entry,
%the stuff in {} is the title that will appear on the first page of the
%table.
\caption{ 89 unknown objects observed with the INT in Feb 2012 in program NEA fields. 
We include approximate orbital data fitted with FIND\_ORB software based on our 1-night datasets.} 
\\
%This is the header for the first page of the table...
\hline \hline \\
   \multicolumn{1}{c}{Acronym} & 
   \multicolumn{1}{c}{$a$ (AU)} &
   \multicolumn{1}{c}{$e$} &
   \multicolumn{1}{c}{$i$ ($\degree$)} &
   \multicolumn{1}{c}{MOID (AU)} &
   \multicolumn{1}{c}{$H$} &
   \multicolumn{1}{c}{Nr pos} &
   \multicolumn{1}{c}{Arc} &
   \multicolumn{1}{c}{$\sigma$ ($^{\prime\prime}$)} &
   \multicolumn{1}{c}{$R$} &
   \multicolumn{1}{c}{$\beta$ ($\degree$)} &
   \multicolumn{1}{c}{$\epsilon$ ($\degree$)} &
   \multicolumn{1}{p{2cm}}{$\mu$ ($^{\prime\prime}$/min)} \\
\\ \hline \hline \\
\endfirsthead

%This is the header for the remaining page(s) of the table...
\multicolumn{13}{c}{{\tablename} \thetable{} (continued) -- 89 unknown objects observed with the INT in Feb 2012 in the opposition mini-survey fields. } \\
\\  \hline \hline \\
   \multicolumn{1}{c}{Acronym} & 
   \multicolumn{1}{c}{$a$ (AU)} &
   \multicolumn{1}{c}{$e$} &
   \multicolumn{1}{c}{$i$ ($\degree$)} &
   \multicolumn{1}{c}{MOID (AU)} &
   \multicolumn{1}{c}{$H$} &
   \multicolumn{1}{c}{Nr pos} &
   \multicolumn{1}{c}{Arc} &
   \multicolumn{1}{c}{$\sigma$ ($^{\prime\prime}$)} &
   \multicolumn{1}{c}{$R$} &
   \multicolumn{1}{c}{$\beta$ ($\degree$)} &
   \multicolumn{1}{c}{$\epsilon$ ($\degree$)} &
   \multicolumn{1}{p{2cm}}{$\mu$ ($^{\prime\prime}$/min)} \\
\\ \hline \hline \\
\endhead

EBA000	&	2.47	&	0.03	&	1.4	&	1.39	&	15.8	&	3	&	1n	&	0.21	&	20.3	&	-1	&	272	&	0.59	\\
EBA002	&	2.68	&	0.15	&	1.0	&	1.30	&	16.0	&	3	&	1n	&	0.09	&	20.1	&	-1	&	272	&	0.74	\\
EBA003	&	2.73	&	0.35	&	9.3	&	0.76	&	14.5	&	4	&	1n	&	0.60	&	20.5	&	-1	&	271	&	0.35	\\
EBA004	&	2.74	&	0.04	&	7.4	&	1.61	&	15.4	&	7	&	1n	&	0.39	&	20.3	&	-1	&	272	&	0.48	\\
EBA005	&	2.65	&	0.08	&	1.5	&	1.42	&	15.2	&	7	&	1n	&	0.39	&	20.2	&	-1	&	272	&	0.46	\\
EBA006	&	3.18	&	0.10	&	1.3	&	1.86	&	13.6	&	7	&	1n	&	0.32	&	18.6	&	-1	&	272	&	0.50	\\
EBA011	&	2.47	&	0.55	&	7.9	&	0.11	&	17.7	&	7	&	1n	&	0.04	&	18.8	&	-15	&	258	&	0.66	\\
EBA012	&	1.06	&	0.11	&	2.3	&	0.01	&	24.5	&	7	&	1n	&	0.18	&	21.0	&	-15	&	258	&	2.13	\\
\end{longtable}
\end{center}

%%\end{document}

%%\documentclass[preprint,authoryear,12pt]{elsarticle}
%%\usepackage{longtable}

%%\begin{document}
%%\setlength{\topmargin}{-0.8in}

% ______________________________

% APPENDIX TABLES
% ______________________________

% ESO/MPG DATA 
%__________

\newpage
%%\appendix
%%\section{Unknown objects observed with the INT during 2011 runs in program NEA fields. } 

\scriptsize
\begin{center}

\begin{longtable}{llrrrrrrrrrrp{2cm}}
%%\begin{longtable}{*{13}{>{\centering\arraybackslash}p{0.85cm}}}
%%\begin{longtable}{p{1.5cm}p{0.85cm}p{0.85cm}p{0.85cm}p{0.85cm}p{0.85cm}p{0.85cm}p{0.85cm}p{0.85cm}p{0.85cm}p{0.85cm}p{0.85cm}p{0.85cm}}

%Here is the caption, the stuff in [] is the table of contents entry,
%the stuff in {} is the title that will appear on the first page of the
%table.
\caption{ 75 unknown objects observed with the INT during 2011 runs in program NEA fields. 
We include approximate orbital data fitted with FIND\_ORB software based on our 1-night datasets.} 
\\
%This is the header for the first page of the table...
\hline \hline \\
   \multicolumn{1}{c}{Acronym} & 
   \multicolumn{1}{c}{$a$ (AU)} &
   \multicolumn{1}{c}{$e$} &
   \multicolumn{1}{c}{$i$ ($\degree$)} &
   \multicolumn{1}{c}{MOID (AU)} &
   \multicolumn{1}{c}{$H$} &
   \multicolumn{1}{c}{Nr pos} &
   \multicolumn{1}{c}{Arc} &
   \multicolumn{1}{c}{$\sigma$ ($^{\prime\prime}$)} &
   \multicolumn{1}{c}{$R$} &
   \multicolumn{1}{c}{$\beta$ ($\degree$)} &
   \multicolumn{1}{c}{$\epsilon$ ($\degree$)} &
   \multicolumn{1}{p{2cm}}{$\mu$ ($^{\prime\prime}$/min)} \\
\\ \hline \hline \\
\endfirsthead

%This is the header for the remaining page(s) of the table...
\multicolumn{13}{c}{{\tablename} \thetable{} (continued) -- 75 unknown objects observed with the INT during 2011 in program NEA fields. } \\
\\  \hline \hline \\
   \multicolumn{1}{c}{Acronym} & 
   \multicolumn{1}{c}{$a$ (AU)} &
   \multicolumn{1}{c}{$e$} &
   \multicolumn{1}{c}{$i$ ($\degree$)} &
   \multicolumn{1}{c}{MOID (AU)} &
   \multicolumn{1}{c}{$H$} &
   \multicolumn{1}{c}{Nr pos} &
   \multicolumn{1}{c}{Arc} &
   \multicolumn{1}{c}{$\sigma$ ($^{\prime\prime}$)} &
   \multicolumn{1}{c}{$R$} &
   \multicolumn{1}{c}{$\beta$ ($\degree$)} &
   \multicolumn{1}{c}{$\epsilon$ ($\degree$)} &
   \multicolumn{1}{p{2cm}}{$\mu$ ($^{\prime\prime}$/min)} \\
\\ \hline \hline \\
\endhead

KVT010	&	2.36	&	0.51	&	3.1	&	0.17	&	21.1	&	6	&	1n	&	0.06	&	21.4	&	-7	&	127	&	0.24	\\
KVT001	&	2.35	&	0.12	&	3.4	&	1.09	&	18.1	&	7	&	1n	&	0.03	&	21.1	&	-5	&	119	&	0.29	\\
KVT002	&	2.56	&	0.01	&	3.9	&	1.54	&	17.9	&	7	&	1n	&	0.10	&	21.9	&	-5	&	119	&	0.14	\\
KVT003	&	2.23	&	0.03	&	5.2	&	1.16	&	18.6	&	7	&	1n	&	0.03	&	22.1	&	-5	&	120	&	0.14	\\
KVT004	&	2.66	&	0.29	&	4.6	&	0.91	&	19.2	&	7	&	1n	&	0.07	&	21.7	&	-5	&	119	&	0.45	\\
KVT005	&	2.30	&	0.11	&	6.1	&	1.05	&	18.1	&	6	&	1n	&	0.09	&	21.9	&	-7	&	127	&	0.17	\\
KVT006	&	2.66	&	0.07	&	14.1	&	1.47	&	17.3	&	6	&	1n	&	0.06	&	21.6	&	-7	&	127	&	0.26	\\
KVT007	&	2.39	&	0.06	&	9.7	&	1.22	&	17.4	&	6	&	1n	&	0.05	&	21.2	&	-7	&	127	&	0.24	\\
\end{longtable}
\end{center}

%%\end{document}

%%\documentclass[preprint,authoryear,12pt]{elsarticle}
%%\usepackage{longtable}

%%\begin{document}
%%\setlength{\topmargin}{-0.8in}

% ______________________________

% APPENDIX TABLES
% ______________________________

% ESO/MPG DATA 
%__________

\newpage
%%\appendix
%%\section{Unknown objects observed with Blanco telescope during one 2011 night in program NEA fields. } 

\scriptsize
\begin{center}
\begin{longtable}{llrrrrrrrrrrp{2cm}}
%%\begin{longtable}{*{13}{>{\centering\arraybackslash}p{0.85cm}}}
%%\begin{longtable}{p{1.5cm}p{0.85cm}p{0.85cm}p{0.85cm}p{0.85cm}p{0.85cm}p{0.85cm}p{0.85cm}p{0.85cm}p{0.85cm}p{0.85cm}p{0.85cm}p{0.85cm}}

%Here is the caption, the stuff in [] is the table of contents entry,
%the stuff in {} is the title that will appear on the first page of the
%table.
\caption{ 196 unknown objects observed with Blanco
during one night in 2011 in program NEA fields. 
We include preliminary orbits fitted with FIND\_ORB.} 
\\
%This is the header for the first page of the table...
\hline \hline \\
   \multicolumn{1}{c}{Acronym} & 
   \multicolumn{1}{c}{$a$ (AU)} &
   \multicolumn{1}{c}{$e$} &
   \multicolumn{1}{c}{$i$ ($\degree$)} &
   \multicolumn{1}{c}{MOID (AU)} &
   \multicolumn{1}{c}{$H$} &
   \multicolumn{1}{c}{Nr pos} &
   \multicolumn{1}{c}{Arc} &
   \multicolumn{1}{c}{$\sigma$ ($^{\prime\prime}$)} &
   \multicolumn{1}{c}{$R$} &
   \multicolumn{1}{c}{$\beta$ ($\degree$)} &
   \multicolumn{1}{c}{$\epsilon$ ($\degree$)} &
   \multicolumn{1}{p{2cm}}{$\mu$ ($^{\prime\prime}$/min)} \\
\\ \hline \hline \\
\endfirsthead

%This is the header for the remaining page(s) of the table...
\multicolumn{13}{c}{{\tablename} \thetable{} (continued) -- 196 unknown objects observed with Blanco during one night in 2011 in program NEA fields. } \\
\\  \hline \hline \\
   \multicolumn{1}{c}{Acronym} & 
   \multicolumn{1}{c}{$a$ (AU)} &
   \multicolumn{1}{c}{$e$} &
   \multicolumn{1}{c}{$i$ ($\degree$)} &
   \multicolumn{1}{c}{MOID (AU)} &
   \multicolumn{1}{c}{$H$} &
   \multicolumn{1}{c}{Nr pos} &
   \multicolumn{1}{c}{Arc} &
   \multicolumn{1}{c}{$\sigma$ ($^{\prime\prime}$)} &
   \multicolumn{1}{c}{$R$} &
   \multicolumn{1}{c}{$\beta$ ($\degree$)} &
   \multicolumn{1}{c}{$\epsilon$ ($\degree$)} &
   \multicolumn{1}{p{2cm}}{$\mu$ ($^{\prime\prime}$/min)} \\
\\ \hline \hline \\
\endhead

PCPV010	&	2.39	&	0.19	&	2.0	&	0.92	&	17.2	&	6	&	1n	&	0.23	&	22.2	&	2	&	277	&	0.54	\\
PCPV011	&	2.49	&	0.02	&	5.0	&	1.42	&	17.3	&	5	&	1n	&	0.61	&	21.9	&	2	&	277	&	0.69	\\
PCPV012	&	2.88	&	0.01	&	13.7	&	1.85	&	17.2	&	3	&	1n	&	0.29	&	22.2	&	2	&	277	&	0.59	\\
PCSV001	&	2.56	&	0.11	&	15.1	&	1.34	&	17.3	&	6	&	1n	&	0.05	&	21.6	&	-20	&	127	&	0.28	\\
PCSV002	&	2.31	&	0.10	&	12.4	&	1.10	&	17.2	&	6	&	1n	&	0.04	&	20.0	&	-20	&	128	&	0.35	\\
PCSV003	&	2.91	&	0.55	&	10.1	&	0.31	&	21.4	&	6	&	1n	&	0.55	&	22.4	&	-20	&	128	&	0.41	\\
PCSV004	&	2.71	&	0.16	&	14.4	&	1.32	&	17.6	&	6	&	1n	&	0.09	&	20.9	&	-20	&	127	&	0.23	\\
PCSV005	&	2.39	&	0.04	&	8.5	&	1.27	&	17.1	&	8	&	1n	&	0.27	&	20.9	&	7	&	101	&	0.59	\\
\end{longtable}
\end{center}

%%\end{document}

%%\documentclass[preprint,authoryear,12pt]{elsarticle}
%%\usepackage{longtable}

%%\begin{document}

% ______________________________

% APPENDIX TABLES
% ______________________________

% ESO/MPG DATA 
%__________

\newpage
%%\appendix

%%\addcontentsline{toc}{chapter}{Appendices}
%%\chapter{Appendix A: Program NEA objects observed with Blanco, INT, WHT, OHP 1.2m, Pic du Midi 1m, Tautenburg 2m, Bonn AIfA 0.5m, Galati 0.4m and Urseanu 0.3m telescopes}

%%\section{TITLU 1}
%%Program NEA objects observed with Blanco, INT, WHT, OHP 1.2m, Pic du Midi 1m, Tautenburg 2m, Bonn AIfA 0.5m, Galati 0.4m and Urseanu 0.3m telescopes} 

\scriptsize
\begin{center}
\begin{longtable}{lrrrrrrrrrrr} 
%Here is the caption, the stuff in [] is the table of contents entry,
%the stuff in {} is the title that will appear on the first page of the
%table.
\caption{ 60 {\it NEO candidates} matching at least one of the three NEO selection methods: 
fit the $\epsilon-\mu$ model, have a good score (greater than $10\%$)
in the ``NEO Rating'' developed by the MPC, or have small calculated MOID
(less than 0.3 AU). We list in bold 18 {\it best NEO candidates} defined to match at least two of the three 
criteria. The object ESU031 marked with $+$ turned out to be the recently
discovered NEA 2012~DC28, and fitted all three NEO 
selection methods well. }
\\
%This is the header for the first page of the table...
\hline \hline \\
   \multicolumn{1}{c}{Acronym} &
   \multicolumn{1}{c}{$\mu$ ($^{\prime\prime}$/min)} &
   \multicolumn{1}{c}{$\epsilon$ ($^\circ$)} &
   \multicolumn{1}{c}{$R$} &
   \multicolumn{1}{c}{MOID (AU)} &
   \multicolumn{1}{c}{$a$ (AU)} &
   \multicolumn{1}{c}{$e$} &
   \multicolumn{1}{c}{$i$ ($\degree$)} &
   \multicolumn{1}{c}{Pos} &
   \multicolumn{1}{c}{$\sigma$ ($^{\prime\prime}$)} &
   \multicolumn{1}{c}{MPC rating} &
   \multicolumn{1}{c}{Model} \\
\\ \hline \hline \\
\endfirsthead

%This is the header for the remaining page(s) of the table...
\multicolumn{10}{c}{{\tablename} \thetable{} (continued) -- 60 NEO candidates and 18 best NEO candidates. } \\
\\  \hline \hline \\
   \multicolumn{1}{c}{Acronym} &
   \multicolumn{1}{c}{$\mu$ ($^{\prime\prime}$/min)} &
   \multicolumn{1}{c}{$\epsilon$ ($^\circ$)} &
   \multicolumn{1}{c}{$R$} &
   \multicolumn{1}{c}{MOID (AU)} &
   \multicolumn{1}{c}{$a$ (AU)} &
   \multicolumn{1}{c}{$e$} &
   \multicolumn{1}{c}{$i$ ($\degree$)} &
   \multicolumn{1}{c}{Pos} &
   \multicolumn{1}{c}{$\sigma$ ($^{\prime\prime}$)} &
   \multicolumn{1}{c}{MPC rate} &
   \multicolumn{1}{c}{Model} \\
\\ \hline \hline \\
\endhead

\multicolumn{12}{c}{INT NEA fields: } \\
\noalign{\smallskip}\noalign{\smallskip}

{\bf EBA012}    &  2.13  &   258  &  21.0  &  {\bf 0.01}  &  1.06  &  0.11  &   2.3  &    7   &   0.18  &  {\bf 100}   &  {\bf Best}  \\
EBA055          &  0.43  &   196  &  21.1  &        0.01  &  1.06  &  0.07  &   0.3  &    5   &   0.20  &          1   &        Bad   \\
EBA067          &  0.47  &   196  &  21.1  &        0.01  &  1.19  &  0.22  &   0.6  &    5   &   0.30  &          0   &        Bad   \\
EBA120          &  0.90  &   178  &  21.1  &        0.78  &  1.94  &  0.10  &  20.9  &    5   &   0.09  &         20   &        Close \\
EPA014          &  1.51  &   284  &  21.2  &        0.68  &  2.89  &  0.44  &  11.5  &    5   &   0.30  &         61   &        Close \\
{\bf EPA015}    &  3.33  &   261  &  19.6  &  {\bf 0.09}  &  1.81  &  0.43  &  22.2  &    6   &   0.49  &  {\bf 100}   &  {\bf Best}  \\
{\bf VFLLP01}   &  1.74  &   261  &  21.7  &  {\bf 0.01}  &  4.72  &  0.79  &   2.2  &    7   &   0.12  &  {\bf  88}   &  {\bf Best}  \\
VKF006          &  0.32  &   160  &  21.2  &        0.45  &  2.29  &  0.38  &   6.9  &    7   &   0.04  &         13   &        Bad   \\
{\bf VKF008}    &  1.13  &   160  &  20.8  &        0.77  &  1.87  &  0.07  &  19.4  &    7   &   0.10  &  {\bf  22}   &  {\bf Best}  \\
VKF012          &  0.31  &   160  &  20.9  &        0.45  &  2.30  &  0.38  &   6.8  &    7   &   0.03  &         11   &        Bad   \\
VKF030          &  0.35  &   186  &  20.8  &        0.53  &  2.34  &  0.35  &   0.3  &    6   &   0.10  &         22   &        Bad   \\
{\bf VKF034}    &  1.22  &   186  &  19.7  &        0.91  &  1.98  &  0.04  &  23.4  &    7   &   0.08  &  {\bf  17}   &  {\bf Best}  \\
VSPK003         &  0.59  &   212  &  21.1  &        0.38  &  2.43  &  0.44  &   6.6  &    7   &   0.13  &          2   &        Good  \\

\noalign{\smallskip}\noalign{\smallskip}
\hline
\noalign{\smallskip}\noalign{\smallskip}
\multicolumn{12}{c}{INT opposition fields: } \\
\noalign{\smallskip}\noalign{\smallskip}

{\bf EBA023}    &  1.11  &   184  &  21.7  &  {\bf 0.01}  &  1.04  &  0.04  &   0.4  &    5   &   0.27  &          8   &  {\bf Good}  \\
EBA024          &  0.80  &   184  &  20.4  &        0.01  &  1.03  &  0.11  &   0.2  &    4   &   0.38  &          1   &       Close  \\
EBA074          &  0.93  &   186  &  21.5  &        0.31  &  1.39  &  0.06  &   0.5  &    3   &   0.27  &          4   &       Close  \\
EBA163          &  0.54  &   184  &  20.7  &        0.01  &  1.31  &  0.39  &   0.2  &    5   &   0.05  &          0   &       Bad    \\
ELA004          &  0.65  &   186  &  20.0  &        0.01  &  1.14  &  0.23  &   0.5  &    5   &   0.27  &          0   &       Bad    \\
ELA014          &  0.54  &   186  &  20.8  &        0.01  &  1.04  &  0.19  &   0.0  &    5   &   0.07  &          0   &       Bad    \\
ELA070          &  0.99  &   182  &  21.0  &        0.90  &  1.95  &  0.03  &  15.6  &    3   &   0.18  &          5   &       Close  \\
EPA109          &  0.64  &   185  &  21.5  &        0.01  &  1.02  &  0.11  &   0.2  &    5   &   0.07  &          0   &       Bad    \\
{\bf EPA143}    &  1.13  &   182  &  20.8  &        0.75  &  1.88  &  0.07  &  16.1  &    5   &   0.09  &  {\bf  11}   &  {\bf Good}  \\
{\bf EPA190}    &  1.05  &   180  &  20.8  &        0.78  &  1.93  &  0.07  &  21.3  &    5   &   0.17  &  {\bf  15}   &  {\bf Good}  \\
{\bf EPO031}    &  1.04  &   182  &  19.6  &  {\bf 0.01}  &  1.06  &  0.14  &   0.4  &    5   &   0.20  &  {\bf  19}   &  {\bf Good}  \\
EPO078          &  0.73  &   186  &  20.3  &        0.01  &  1.50  &  0.47  &   0.4  &    5   &   0.02  &          1   &       Bad    \\
EPO026          &  0.61  &   185  &  21.0  &        0.01  &  1.14  &  0.27  &   0.5  &    4   &   0.32  &          0   &       Bad    \\
{\bf EPO065}    &  1.08  &   186  &  20.6  &        0.82  &  1.95  &  0.06  &  22.8  &    4   &   0.10  &  {\bf  17}   &  {\bf Good}  \\
ETU018          &  0.55  &   183  &  20.9  &        0.01  &  1.02  &  0.03  &   0.2  &    5   &   0.13  &          1   &       Bad    \\
ETU091          &  0.63  &   183  &  21.5  &        0.15  &  1.30  &  0.10  &   0.2  &    5   &   0.10  &          0   &       Bad    \\
ETU108          &  0.47  &   183  &  20.7  &        0.03  &  1.04  &  0.04  &   0.3  &    5   &   0.03  &          0   &       Bad    \\
ETU125          &  0.61  &   186  &  20.7  &        0.01  &  1.02  &  0.05  &   0.3  &    5   &   0.47  &          0   &       Bad    \\
{\bf ESU031 + } & 10.32  &   183  &  20.5  &  {\bf 0.01}  &  1.02  &  0.01  &   6.7  &    3   &   0.56  &  {\bf 100}   &  {\bf Best}  \\
ESU096          &  0.53  &   187  &  21.5  &        0.01  &  1.07  &  0.38  &   0.1  &    4   &   0.06  &          0   &       Bad    \\
ESU110          &  0.57  &   187  &  21.2  &        0.01  &  1.19  &  0.27  &   0.4  &    5   &   0.11  &          0   &       Bad    \\
ESU166          &  0.81  &   186  &  20.8  &        0.01  &  1.02  &  0.08  &   0.2  &    5   &   0.29  &          2   &       Close  \\
ESU174          &  0.57  &   187  &  20.7  &        0.01  &  1.07  &  0.08  &   0.3  &    5   &   0.09  &          0   &       Bad    \\
ESU181          &  0.49  &   187  &  21.3  &        0.01  &  1.35  &  0.40  &   0.7  &    5   &   0.17  &          0   &       Bad    \\

\noalign{\smallskip}\noalign{\smallskip}
\hline
\noalign{\smallskip}\noalign{\smallskip}
\multicolumn{12}{c}{Blanco NEA fields: } \\
\noalign{\smallskip}\noalign{\smallskip}

PCSV007         &  0.74  &   102  &  22.4  &        0.01  &  1.14  &  0.11  &   0.2  &    7   &   0.19  &          2   &       Bad    \\
PCSV026         &  1.05  &    79  &  21.3  &        0.01  &  1.00  &  0.01  &   0.3  &    8   &   0.06  &          4   &       Bad    \\
PCSV027         &  0.90  &    79  &  22.6  &        0.01  &  1.01  &  0.02  &   0.2  &    8   &   0.15  &          1   &       Bad    \\
PCTV007         &  1.07  &   293  &  22.2  &        0.03  &  0.97  &  0.18  &   2.0  &    6   &   0.07  &          1   &       Bad    \\
PCTV086         &  0.41  &   104  &  23.2  &        0.01  &  1.31  &  0.22  &   0.9  &    6   &   0.25  &          0   &       Bad    \\
{\bf PCTV024}   &  0.66  &   169  &  18.3  &  {\bf 0.28}  &  2.51  &  0.48  &   0.7  &    5   &   0.12  &  {\bf 100}   &       Bad    \\
{\bf PCTV026}   &  0.48  &   169  &  19.9  &  {\bf 0.28}  &  2.23  &  0.42  &   1.1  &    3   &   0.08  &  {\bf 100}   &       Bad    \\
PCTV028         &  0.33  &   232  &  21.5  &        0.03  &  1.13  &  0.08  &   0.7  &    8   &   0.10  &          7   &       Bad    \\
PCTV098         &  0.39  &   104  &  22.9  &        1.37  &  2.45  &  0.04  &   6.1  &    8   &   0.20  &          0   &       Bad    \\
{\bf PCTVb50}   &  1.07  &    91  &  20.4  &  {\bf 0.03}  &  1.15  &  0.17  &   2.3  &    6   &   0.06  &  {\bf  10}   &       Bad    \\
PCTVb58         &  0.95  &    91  &  22.6  &        0.01  &  1.31  &  0.22  &   0.4  &    8   &   0.21  &          1   &       Bad    \\
PCTVb63         &  0.25  &    91  &  20.8  &        0.43  &  2.54  &  0.44  &   3.8  &    8   &   0.07  &         22   &       Bad    \\
PCTVb68         &  1.13  &    91  &  22.1  &        0.01  &  1.02  &  0.00  &   0.2  &    6   &   0.26  &          3   &       Bad    \\
PCTVb69         &  1.00  &    91  &  22.6  &        0.01  &  1.01  &  0.00  &   0.1  &    8   &   0.22  &          5   &       Bad    \\
PCTVb71         &  0.64  &    91  &  22.8  &        0.02  &  1.49  &  0.41  &   1.9  &    7   &   0.14  &          0   &       Bad    \\
PCV013          &  0.20  &   231  &  20.0  &        0.47  &  2.56  &  0.43  &  15.1  &    7   &   0.66  &         77   &       Bad    \\
{\bf PCV023}    &  1.78  &   269  &  20.7  &  {\bf 0.15}  &  2.11  &  0.45  &   5.1  &    4   &   0.12  &  {\bf  80}   &  {\bf Good}  \\
{\bf PCVP005}   &  0.62  &   248  &  22.4  &  {\bf 0.25}  &  1.91  &  0.61  &  22.5  &    3   &   0.10  &  {\bf 100}   &       Bad    \\
{\bf PCVP007}   &  0.34  &   276  &  21.9  &  {\bf 0.14}  &  3.20  &  0.84  &  17.6  &    4   &   0.22  &  {\bf 100}   &       Bad    \\
PCVP013         &  0.63  &   189  &  19.1  &        1.16  &  2.56  &  0.16  &  14.2  &    6   &   0.17  &         10   &       Bad    \\
PCVP023         &  0.87  &   189  &  21.6  &        0.73  &  1.97  &  0.17  &  19.2  &    3   &   0.20  &         23   &       Close  \\
PCVS010         &  0.39  &   269  &  21.9  &        2.20  &  3.21  &  0.01  &   4.7  &    6   &   0.18  &          1   &       Bad    \\
{\bf PCVS024}   &  0.31  &   296  &  22.2  &  {\bf 0.05}  &  2.28  &  0.93  &   5.2  &    4   &   0.16  &  {\bf  19}   &       Bad    \\

\\ \hline \hline \\

\end{longtable}
\end{center}

% _________

%%\end{document}